\documentclass[reprint,
superscriptaddress,
amsmath,amssymb,
aps,
pra,
]{revtex4-2}

\usepackage{graphicx}
\usepackage{dcolumn}
\usepackage{bm}

\begin{document}

\title{When Do Quantum Correlations Matter for Quantum Annealing?
}


\author{Rongfeng Xie}
\affiliation{School of Physics and Astronomy, University of Minnesota, Minneapolis, Minnesota 55455, USA}

\author{Evgenii Ievlev}
\altaffiliation[Present address: ]{School of Science and Engineering, The Chinese University of Hong Kong (Shenzhen), Shenzhen, 518172, China}
\affiliation{School of Physics and Astronomy, University of Minnesota, Minneapolis, Minnesota 55455, USA}
\affiliation{
W. I. Fine Theoretical Institute, University of Minnesota, Minneapolis, Minnesota 55455, USA}

\author{Alex Kamenev}
\affiliation{School of Physics and Astronomy, University of Minnesota, Minneapolis, Minnesota 55455, USA}
\affiliation{
W. I. Fine Theoretical Institute, University of Minnesota, Minneapolis, Minnesota 55455, USA}

\date{\today}

\begin{abstract}

Whether quantum correlations give non-adiabatic annealing a scaling advantage remains an open question. We address it by comparing quantum annealing (QA) with quantum-inspired annealing (QIA), a classical protocol that retains the Hamiltonian structure and annealing schedule of QA while excluding nonlocal quantum correlations. Across models varying in dimensionality, disorder, and frustration, we find that the difference in scaling between the two protocols is strongly model dependent. QA shows a clear advantage in clean, low-dimensional systems, but the advantage diminishes with increasing dimensionality and becomes negligible in three or more dimensions. Disorder and frustration further narrow the difference. These results identify where quantum correlations have a qualitative effect on annealing scaling and provide a classical baseline for assessing their role.

\end{abstract}


\maketitle

\section{Introduction}

Quantum annealing (QA) seeks to solve optimization problems by encoding a cost function in an Ising-like Hamiltonian and evolving the system with a time-dependent protocol~\cite{Albash_2018,brooke_1999,farhi_2000,kadowaki_1998,morita_2008,santoro_2002}. The system begins in the ground state of a transverse-field driver, after which the Hamiltonian is interpolated toward a problem Hamiltonian whose ground state represents the desired solution. Quantum fluctuations may help the system traverse barriers that impede classical dynamics, making QA a potential route to improved optimization performance~\cite{Ray1989,Das2008,Katzgraber2014,Heim2015,Denchev2016,Mandra2016,Crosson2016}. This prospect has motivated extensive theoretical work and experiments on programmable annealing platforms~\cite{Johnson2011,Dickson2013,Lanting2014,Boixo2014,Ronnow2014,Harris2018,King2022,King2023,zhang2024cyclic,zhang2025computational}.

In principle, a sufficiently slow anneal ends up in the ground state of the final Hamiltonian by the adiabatic theorem~\cite{Albash_2018,farhi_2000,morita_2008}. The required annealing time, however, is controlled by the minimum gap in the adiabatic spectrum, which shrinks exponentially with system size for difficult problems. Large-scale devices therefore operate in a non-adiabatic regime in which excitations are unavoidable. This is the regime studied here: we ask how the density and energy of the remaining excitations scale with annealing time, well before the finite-size adiabatic limit is reached.

The nature and the very existence of a quantum advantage in this regime remains unclear~\cite{Ronnow2014,Katzgraber2014,Heim2015}. Observed QA behavior can reflect entanglement and other quantum correlations, but it can also arise from the open system effects, interpolation schedule, critical slowing down, or slow relaxation in disordered and glassy landscapes~\cite{Das2008,Albash_2018,Amin2015,Mandra2016}. Thus, a strong performance by a quantum annealer does not by itself show that quantum correlations are decisive. The distinction is especially important given the competitive performance of quantum-inspired classical algorithms on Ising-like optimization problems~\cite{arai2024,xiao2026,tao2026}. The sharper question is therefore: when does QA owe its outcome  specifically to quantumness, and when is the dominant physics already captured by a closely matched classical dynamics?

To isolate that role, we use a classical baseline related to semiclassical models previously employed in QA benchmarks~\cite{Shin2014,Subires2022,Bowles2022,Misra2023}, sometimes referred to as a quantum-inspired annealing (QIA). QIA uses the same time-dependent Hamiltonian and schedule as QA but restricts the state to a product-state manifold. It therefore retains 3D mean-field spin precession while excluding entanglement and other nonlocal quantum correlations. Comparing QA with QIA helps identify the role of these correlations in annealing dynamics and residual-energy scaling. We use D-Wave Advantage and Advantage2 devices as proxies for QA and compare their scaling with classically simulated QIA dynamics.    

The focus of this work is not on establishing an end-to-end computational speedup, but on identifying when quantum correlations change the leading {\em scaling} of the annealing dynamics. The comparison reveals several distinct outcomes. In the clean low-dimensional transverse-field Ising models, QA and QIA belong to different dynamical universality classes and QA exhibits a clear scaling advantage. The fact that the experimental QA scaling agrees well with the analytical expectations (distinct from QIA ones) indicates the presence of non-local quantum correlations in the experimental setup \cite{King2022}.  However, the  quantum scaling advantage gradually decreases when the dimensionality increases, becoming very small already in 3D. Finally, in the infinite-dimensional Lipkin-Meshkov-Glick model, the two protocols are shown to exhibit exactly the same  scaling. Similarly, in random-bond spin chains and in short-range 2D and 3D spin-glass models, our data show essentially identical scaling in QA vs. QIA protocols over the accessible time window. 

The rest of the paper is organized as follows: Section~II derives QIA equations of motion as a product-state truncation of the full quantum dynamics. Section~III compares QA and QIA in clean, disordered, and frustrated models, and Sec.~IV discusses the implications and limitations of the comparison.

\section{Quantum-Inspired Annealing}
Consider optimization problems encoded in an Ising-like cost function on a graph
$G=(V,E)$ with $N=|V|$ vertices,
\begin{equation}
E(\{s_i\})=
-\sum_{(i,j)\in E} J_{ij}s_is_j
-\sum_{i=1}^N h_i s_i,
\qquad s_i\in\{-1,+1\}.
\label{eq:ising_cost}
\end{equation}
For simplicity, we set $h_i=0$ throughout this work. The corresponding QA protocol is defined by the time-dependent Hamiltonian
\begin{equation}
H(t)=
-A(t)\sum_{i=1}^N \sigma_i^x
-B(t)\sum_{(i,j)\in E}J_{ij}\sigma_i^z\sigma_j^z,
\label{eq:qa_hamiltonian}
\end{equation}
where $\sigma_i^\alpha$ ($\alpha=x,y,z$) denote the Pauli operators acting on spin $i$, and $A(t)$ and $B(t)$ interpolate between a transverse-field driver and the classical Ising problem Hamiltonian. Unless otherwise specified, we use a dimensionless linear schedule,
\begin{equation}
A(t)=1-f(t),\qquad B(t)=f(t),\qquad f(t)=\frac{t}{\tau_{\rm ann}},
\label{eq:linear_schedule}
\end{equation}
with total annealing time $\tau_{\rm ann}$.

To isolate the role of quantum correlations while preserving the Hamiltonian structure and annealing schedule of QA, we use a closely matched classical baseline, which we refer to as quantum-inspired annealing (QIA). Closely related approaches have been considered previously in the context of quantum annealing and optimization~\cite{Misra2023}. The basic idea is to retain the same time-dependent Hamiltonian and schedule as in QA, but to restrict the dynamics to a product-state manifold. In this way, QIA preserves the self-consistent spin dynamics generated by the annealing protocol while excluding entanglement and other nonlocal quantum correlations.

The local magnetizations are defined as
\begin{equation}
m_i^\alpha(t)=\langle \sigma_i^\alpha\rangle
={\rm Tr}\!\left[\rho(t)\sigma_i^\alpha\right],
\qquad \alpha\in\{x,y,z\},
\label{eq:local_magnetization}
\end{equation}
and the connected two-point correlators
\begin{equation}
C_{ij}^{\alpha\beta}
=
\langle \sigma_i^\alpha \sigma_j^\beta\rangle
-
m_i^\alpha m_j^\beta.
\label{eq:connected_correlator}
\end{equation}
Higher-order connected correlators are defined analogously. Under the exact quantum dynamics, governed by the von Neumann equation,
\begin{equation}
    \dot \rho = -i\left[H,\rho \right],
\end{equation}
these quantities satisfy a hierarchy of coupled equations of motion: the evolution of the local magnetizations depends on two-point correlators, the evolution of two-point correlators depends on three-point correlators, and so on.

The defining approximation of QIA is to truncate this hierarchy by neglecting all connected correlators of order two and higher between distinct sites,
\begin{equation}
C_{ij}^{\alpha\beta}=0,\qquad
C_{ijk}^{\alpha\beta\gamma}=0,\qquad \ldots
\label{eq:qia_truncation}
\end{equation}
so that expectation values of operator products factorize as
\begin{equation}
\langle \sigma_i^\alpha \sigma_j^\beta\rangle
\approx
\langle \sigma_i^\alpha\rangle
\langle \sigma_j^\beta\rangle,
\qquad i\neq j.
\label{eq:qia_factorization}
\end{equation}
Equivalently, the many-body state is approximated by a product state at all times. This truncation leaves the annealing Hamiltonian unchanged, but removes the nonlocal quantum correlations that distinguish the fully quantum dynamics.

Under this approximation, each spin evolves in a self-consistent effective longitudinal field generated by the instantaneous magnetizations of the other spins,
\begin{equation}
h_i^{\rm eff}(t)=\sum_{j\in\partial i} J_{ij} m_j^z(t).
\label{eq:effective_field}
\end{equation}
Here, $\partial i$ denotes the set of sites coupled to site $i$. The resulting equations of motion for the local magnetization vectors take the  form of Landau-Lifshits equations
\begin{equation}
\dot{\mathbf m}_i
=
2\,\mathbf m_i \times \mathbf b_i,
\qquad
\mathbf b_i(t)=\bigl(A(t),\,0,\,B(t)h_i^{\rm eff}(t)\bigr),
\label{eq:qia_vector_eom}
\end{equation}
or, in components,
\begin{subequations}
\label{eq:qia_eom}
\begin{align}
\dot m_i^x &= 2B(t)\,h_i^{\rm eff}(t)\,m_i^y,
\label{eq:qia_mx}\\
\dot m_i^y &= 2A(t)\,m_i^z - 2B(t)\,h_i^{\rm eff}(t)\,m_i^x,
\label{eq:qia_my}\\
\dot m_i^z &= -2A(t)\,m_i^y.
\label{eq:qia_mz}
\end{align}
\end{subequations}
These equations describe deterministic nonlinear precession of coupled local spin vectors in a time-dependent self-consistent field. The initial spins are nearly polarized along the positive \(x\) direction, with small independent random angular deviations of at most \(0.1\) radians that seed the instability, as specified in Appendix~\ref{app:qia_equations}. Notice that the deterministic QIA dynamics is very different from Monte-Carlo Glauber or Kawasaki spin-flip evolution, often associated with the classical simulated annealing. 

QIA should therefore be viewed as a closely matched classical baseline for QA: it preserves the problem Hamiltonian, the driver term, and the annealing schedule, but excludes the nonlocal quantum correlations present in the full quantum evolution. Comparing QA and QIA across different models then allows us to determine when such correlations make a leading-order difference to annealing performance, and when the dominant behavior is already captured at the level of self-consistent product-state dynamics. The QIA evolution can be efficiently modeled on conventional computers even for rather large systems, see Appendix \ref{app:qia_equations} for details.

\section{Performance Across Models: Dimensionality, Disorder, and Frustration}

\subsection{Clean transverse-field Ising models in one, two, and three dimensions}
\label{sec:clean_ising}

Clean ferromagnetic Ising models provide a setting for examining the role of spatial dimensionality without introducing disorder or frustration. The comparison includes a one-dimensional chain, a two-dimensional square lattice, and a three-dimensional cubic lattice, with Hamiltonian
\begin{equation}
H(t)=-A(t)\sum_i\sigma_i^x-JB(t)\sum_{\langle ij\rangle_d}\sigma_i^z\sigma_j^z,
\end{equation}
where \(\langle ij\rangle_d\) denotes nearest-neighbor pairs in dimension \(d\), and the uniform ferromagnetic coupling is set to \(J=1\). The systems contain \(N=1000\) spins in one dimension, \(30\times30\) spins in two dimensions, and \(10\times10\times10\) spins in three dimensions, all with open boundary conditions. QA results are obtained experimentally on the D-Wave annealer, whereas QIA results follow from numerical integration of the product-state equations of motion.

Annealing performance across the three geometries is characterized by the final residual energy density,
\begin{equation}
\epsilon_{\mathrm{res}}=\frac{\langle H_{\mathrm{Ising}}\rangle_f-E_0}{N},
\label{eq:eres}
\end{equation}
where \(H_{\mathrm{Ising}}=-J\sum_{\langle ij\rangle_d}\sigma_i^z\sigma_j^z\), \(E_0\) is its ground-state energy, and \(\langle\cdot\rangle_f\) denotes the final-state expectation value. For uniform ferromagnetic couplings, this can also be written as
\begin{equation}
\epsilon_{\mathrm{res}}=\frac{J}{N}\sum_{\langle ij\rangle_d}\left(1-\langle\sigma_i^z\sigma_j^z\rangle_f\right).
\end{equation}
Within the product-state approximation, the two-spin expectation value factorizes as \(\langle\sigma_i^z\sigma_j^z\rangle_f=m_i^zm_j^z\), where \(m_i^z=\langle\sigma_i^z\rangle_f\). For QA, the final-state expectation values are estimated by averaging the corresponding observables over 1000 spin configurations measured in the \(z\) basis. For QIA, each final magnetization configuration is converted into an Ising spin configuration by taking the sign of \(m_i^z\) at each site, and the reported observables are averaged over 100 such configurations.

The Kibble--Zurek (KZ) mechanism provides a framework for relating excitation production during annealing across critical regime. Near a continuous transition, the equilibrium correlation length and characteristic relaxation time obey
\begin{equation}
\xi\sim|A-A_c|^{-\nu},\qquad \tau_{\mathrm{rel}}\sim\xi^z,
\end{equation}
where \(A\) is the instantaneous control parameter, \(A_c\) is its critical value, and \(\nu\) and \(z\) are the correlation-length and dynamical critical exponents. For a passage that is locally linear near the critical point, the KZ mechanism~\cite{Kibble1976,Zurek1985,Zurek1996,Dziarmaga2010,Polkovnikov2011} predicts a characteristic length
\begin{equation}
\hat\xi\sim\tau_{\mathrm{ann}}^{\nu/(1+z\nu)}.
\end{equation}
In regimes where the final residual energy density exhibits algebraic decay, its dependence on annealing time is characterized by the exponent \(\alpha\),
\begin{equation}
\epsilon_{\mathrm{res}}\propto\tau_{\mathrm{ann}}^{-\alpha}.
\end{equation}
Relating \(\alpha\) to the critical exponents requires connecting the characteristic length \(\hat\xi\) to the energy of the excitations remaining at the end of the anneal.

This connection is most direct in the clean one-dimensional chain, where the domain-wall density and residual energy density satisfy
\begin{equation}
\begin{aligned}
n_{\mathrm{dw}}&=\frac{1}{2N}\sum_{\langle ij\rangle_1}\left(1-\langle\sigma_i^z\sigma_j^z\rangle_f\right),\\
\epsilon_{\mathrm{res}}&=2J n_{\mathrm{dw}}.
\end{aligned}
\end{equation}
Each domain wall carries the same energy cost \(2J\). With \(n_{\mathrm{dw}}\sim\hat\xi^{-1}\), the KZ estimate therefore yields \(\alpha=\nu/(1+z\nu)\). For the quantum Ising chain, \(\nu=z=1\), yielding \(\hat\xi_{\mathrm{QA}}\sim\tau_{\mathrm{ann}}^{1/2}\) and
\begin{equation}
n_{\mathrm{dw}}^{\mathrm{QA}},\ \epsilon_{\mathrm{res}}^{\mathrm{QA}}\sim\tau_{\mathrm{ann}}^{-1/2}.
\end{equation}
This result agrees with exact solutions of the quantum chain~\cite{Dziarmaga2005,Zurek2005} and experimental observations of coherent annealing~\cite{Bando2020,King2022}. For QIA, the semiclassical instability analyzed in Appendix~\ref{app:qia_1d_critical} instead has \(\nu_{\mathrm{QIA}}=1/2\) and \(z_{\mathrm{QIA}}=1\). The corresponding KZ estimate is
\begin{equation}
\begin{aligned}
\hat\xi_{\mathrm{QIA}}&\sim\tau_{\mathrm{ann}}^{1/3},\\
n_{\mathrm{dw}}^{\mathrm{QIA}},\ \epsilon_{\mathrm{res}}^{\mathrm{QIA}}&\sim\tau_{\mathrm{ann}}^{-1/3}.
\end{aligned}
\end{equation}
A related one-third exponent arises for underdamped symmetry-breaking dynamics in a classical field model~\cite{LagunaZurek1998}.

\begin{figure*}[t]
\centering
\begin{minipage}[t]{0.32\textwidth}
\centering (a) 1D chain\\[2pt]
\includegraphics[width=\linewidth]{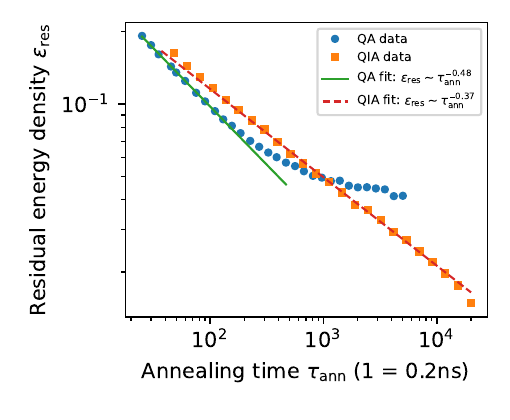}
\end{minipage}\hfill
\begin{minipage}[t]{0.32\textwidth}
\centering (b) 2D square lattice\\[2pt]
\includegraphics[width=\linewidth]{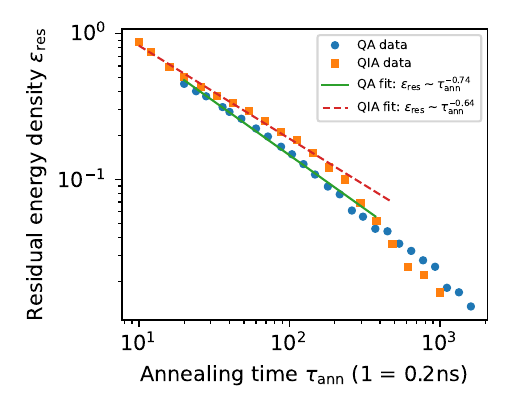}
\end{minipage}\hfill
\begin{minipage}[t]{0.32\textwidth}
\centering (c) 3D cubic lattice\\[2pt]
\includegraphics[width=\linewidth]{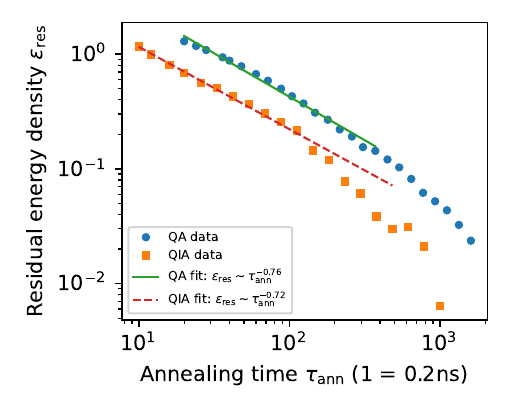}
\end{minipage}
\caption{Residual energy density versus annealing time for clean ferromagnetic
Ising models: (a) a chain with $N=1000$, (b) a $30\times30$ square lattice,
and (c) a $10\times10\times10$ cubic lattice, all with open boundary
conditions. QA data are measured on the D-Wave annealer; QIA data are
obtained by numerical integration. Unless otherwise specified, QA and QIA
results are averaged over $1000$ and $100$ samples, respectively.
One time unit corresponds to $0.2\,\mathrm{ns}$ of hardware annealing time.
Lines show power-law fits
$\epsilon_{\rm res}\propto\tau_{\rm ann}^{-\alpha}$,
with QA/QIA exponents $0.48/0.37$, $0.74/0.64$, and $0.76/0.72$ in
panels (a), (b), and (c), respectively.}
\label{fig:clean_tfim}
\label{fig:1d_tfim}
\end{figure*}

Figure~\ref{fig:clean_tfim}(a) compares these predictions with the experimental QA data and numerical QIA data for the clean one-dimensional Ising chain. The fitted residual-energy exponents are \(\alpha_{\mathrm{QA}}\simeq0.48\) and \(\alpha_{\mathrm{QIA}}\simeq0.37\), close to the respective KZ values \(1/2\) and \(1/3\). Thus, over the fitted regime, QA exhibits a faster decay of residual energy than the product-state baseline. At longer annealing times, the hardware data flatten, consistent with the influence of open-system effects~\cite{King2022,King2023}. The comparison with coherent critical dynamics is therefore restricted to the regime preceding this departure.

The square- and cubic-lattice results are shown in Figs.~\ref{fig:clean_tfim}(b) and \ref{fig:clean_tfim}(c). Over the fitted time windows, the best fits for the residual-energy exponents are
\begin{equation}
(\alpha_{\mathrm{QA}},\alpha_{\mathrm{QIA}})\simeq
\begin{cases}
(0.74,\,0.64),&d=2,\\
(0.76,\,0.72),&d=3.
\end{cases}
\end{equation}
At longer annealing times, the QIA curves bend downward relative to the fitted power laws in both two and three dimensions, reflecting  crossover toward adiabatic behavior in finite-size systems. The corresponding downturn is less pronounced in the experimental QA data, likely due to open-system effects masking  approach to the adiabatic ground-state preparation. The quoted exponents characterize the intermediate-time regime preceding these departures. The separation between the fitted QA and QIA exponents decreases  with increasing dimension and is smallest in the 3D system. 

Previous work has interpreted the residual-energy scaling of the two-dimensional quantum Ising model using a phenomenological ansatz motivated by KZ scaling \(\alpha=d\nu/(1+z\nu)\)~\cite{Weinberg2020}. For the two-dimensional quantum Ising model, \(z=1\) and \(d+z=3\), so the transition belongs to the three-dimensional classical Ising universality class, with \(\nu\simeq0.630\). Substituting these exponents into the ansatz gives \(\alpha\simeq0.773\), close to the fitted QA exponent \(0.74\). The same study reported an experimental exponent \(0.74\pm0.02\). For QIA, the large-\(S\) analysis in Appendix~\ref{app:qia_higher_d} gives \(\nu_{\mathrm{QIA}}=1/2\) and \(z_{\mathrm{QIA}}=1\). Applying the same ansatz results in \(\alpha_{\mathrm{QIA}}=2/3\), also close to the fitted value \(0.64\). While these numerical comparisons are suggestive, they  do not yet establish validity of the ansatz \cite{Weinberg2020} for all $d$. 

For the three-dimensional quantum Ising model, the effective dimension is \(d+z=4\), corresponding to the upper critical dimension of the classical Ising model. The leading critical exponents are therefore \(\nu=1/2\) and \(z=1\), with logarithmic corrections. Such corrections have recently been examined for ramps ending at the critical point~\cite{Naik2026}. The leading exponents coincide with those obtained for QIA in Appendix~\ref{app:qia_higher_d}. Substituting these values into the same ansatz, \(\alpha=d\nu/(1+z\nu)\), brings \(\alpha=1\) for both protocols, appreciably larger than the fitted values \(0.76\) and \(0.72\). Thus, although the observed slopes are close, their values are not explained by  ansatz \cite{Weinberg2020}, and their proximity alone does not establish a common asymptotic scaling law.

The issue at hand is that in \(d>1\), the connection between KZ critical length scale and the final residual energy is complicated by the geometry of extended domain boundaries and by the post-transition coarsening evolution within the ordered phase. Indeed, coarsening through domain growth and interface motion, together with redistribution of energy among excitations, can modify the final residual energy. Studies of isolated two-dimensional quantum Ising dynamics have identified departures from near-critical KZ scaling and additional time scales associated with post-critical ordering~\cite{Schmitt2022,WeinbergXuSandvik2025}. These studies suggest that post-critical evolution may also affect the final residual energy in QIA, although its role in the QIA dynamics remains to be established. To conclude this section: the $d>1$ experimental results demonstrate a decreasing difference between QA and QIA scaling exponents over the observed range of annealing times. 

\subsection{Lipkin-Meshkov-Glick model}

The Lipkin-Meshkov-Glick (LMG) model~\cite{Lipkin1965} provides the next comparison and can be viewed as the infinite-range, or infinite-dimensional, limit of the transverse-field Ising model. This extends our comparison of QA and QIA from finite-dimensional lattices to a fully connected system. The Hamiltonian takes the form
\begin{equation}
H(t)
= - A(t)\sum_i \sigma_i^x 
- \frac{B(t)}{N}\sum_{i<j}\sigma_i^z\sigma_j^z,
\end{equation}
where the \(1/N\) normalization ensures an extensive energy in the thermodynamic limit. For the quantum annealing simulations, we exploit the permutation symmetry of the LMG Hamiltonian. Starting from the fully polarized initial state, the dynamics remains in the maximal-spin sector \(S=N/2\), so that the problem reduces to the evolution of a single collective spin with Hilbert-space dimension \(N+1\), rather than \(2^N\). This allows us to simulate large N systems by direct numerical integration of the time-dependent Schrödinger equation. Despite its infinite-range character, the LMG model exhibits a well-defined quantum phase transition separating a paramagnetic phase from a symmetry-broken ferromagnetic phase~\cite{Botet1982,Xie2024}. Unlike the one-dimensional chain, however, the absence of spatial locality means that concepts such as correlation length, domain walls, and topological defects are no longer well defined. Consequently, the KZ mechanism in its conventional defect-counting form does not directly apply. The residual energy density defined in Eq.~\ref{eq:eres} therefore serves as the measure of annealing performance, which measures the excess energy per spin above the final ground state.

\begin{figure}[t]
    \centering
    \includegraphics[width=0.8\linewidth]{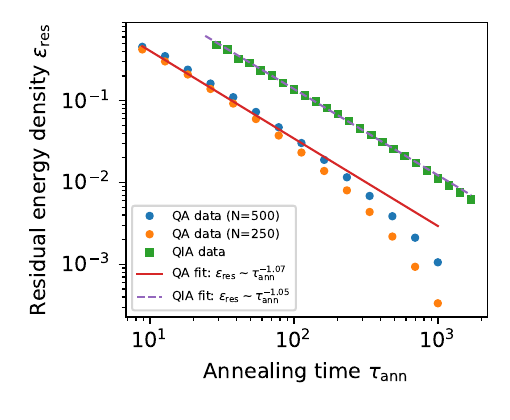}
    \caption{Residual energy density as a function of annealing time for the Lipkin--Meshkov--Glick model. Both QA and QIA data are obtained from numerical simulations, since the all-to-all connectivity of the model is not directly realizable on current D-Wave hardware. The QA results for $N=250$ and $N=500$ collapse onto the same leading behavior, where the downtrend indicates finite-size adiabatic effects in the scaling regime shown. Both protocols exhibit essentially identical algebraic decay, with fitted exponents $\epsilon_{\mathrm{res}} \sim \tau_{\rm ann}^{-1.07}$ for QA and $\epsilon_{\mathrm{res}} \sim \tau_{\rm ann}^{-1.05}$ for QIA, consistent with the expected asymptotic scaling $\epsilon_{\mathrm{res}} \sim \tau_{\rm ann}^{-1}$. This agreement shows that, in the LMG model, quantum correlations do not produce a distinct leading-order advantage over the classical baseline.}
    \label{fig:LMG}
\end{figure}

As shown in Fig.~\ref{fig:LMG}, both QA and QIA exhibit essentially identical scaling behavior of the residual energy density as a function of the total annealing time. This observation can be understood from the all-to-all coupling structure of the model, which reduces the dynamics to that of a single collective spin of length \(S=N/2\) subject to a time-dependent effective field. Both QA and QIA therefore describe coherent or semiclassical precession of this collective degree of freedom, differing only in the presence or absence of quantum correlations. Crucially, quantum correlations do not appear to modify the leading nonadiabatic scaling near the critical point. Moreover, the numerical results are consistent with an approximately inverse-time decay of the residual energy density over the annealing-time range studied,
\begin{equation}
\epsilon_{\rm res}^{\rm QA} \sim \epsilon_{\rm res}^{\rm QIA} \sim \tau_{\rm ann}^{-1}.
\end{equation}
For QIA, this scaling is supported by an analysis of the semiclassical dynamics presented in Appendix~\ref{app:lmg}. A related nondissipative scalar-field model was recently studied in Ref.~\cite{Tyagi2025}, where the same KZ scaling exponent was found for nonadiabatic excitations generated during a sweep through a critical point. The agreement in the leading scaling between QA and QIA should be understood as a thermodynamic-limit result in the nonadiabatic regime. At finite $N$, finite-size quantum effects introduce a characteristic crossover time scale $\tau_N\sim N$~\cite{Xie2024, Li2026}. For $\tau_{\rm ann}\ll\tau_N$, QA follows the same leading semiclassical scaling as QIA, whereas for $\tau_{\rm ann} \gtrsim \tau_N$, finite-size quantum effects become important and the two protocols need no longer exhibit the same dynamics.

The LMG model thus represents a regime in which infinite-dimensional collective dynamics suppresses the quantitative advantage associated with quantum correlations. Despite the presence of a genuine quantum phase transition, the effective reduction to a single collective mode makes the leading scaling insensitive to entanglement and other nonlocal quantum effects. Together with Sec.~\ref{sec:clean_ising}, this model provides a controlled contrast to the clean chain: its collective dynamics yields the same leading scaling for QA and QIA, whereas the short-range higher-dimensional comparisons remain empirical over the accessible time windows. The following sections examine how disorder and frustration further modify this picture.

\subsection{One-dimensional disordered transverse-field Ising model}
The one-dimensional transverse-field random-bond Ising chain is described by
\begin{equation}
H(t) =
-A(t)\sum_i \sigma_i^x
-B(t)\sum_i J_{i,i+1}\sigma_i^z\sigma_{i+1}^z,
\end{equation}
where the ferromagnetic couplings $J_{i,i+1}$ are independent random variables drawn from the uniform distribution $J_{i,i+1}\in [0,1]$, and the annealing schedule is chosen as in the previous sections. This system contains no frustration. Nevertheless, random bonds create weak and strong links along the chain, so different regions respond to the annealing dynamics on widely different time scales. Although the model is not a spin glass in the usual sense, it shares this important qualitative feature with glassy systems~\cite{Derrida1986}. As such, it provides a controlled setting for isolating how disorder affects the role of quantum correlations in annealing dynamics, in the absence of frustration.

The random-bond transverse-field Ising chain has been extensively studied and its phase transition is known to be governed by an infinite-randomness quantum critical point~\cite{Igloi2005,Vojta2006}. The strong-disorder real-space renormalization-group approach was pioneered by Ma, Dasgupta, and collaborators in the study of random quantum spin chains~\cite{Ma1979,Dasgupta1980}, and was subsequently developed for the random transverse-field Ising chain by D. Fisher and later works~\cite{Fisher1995,Igloi2001,Igloi2002}. At the critical point, conventional power-law dynamical scaling is replaced by activated scaling, in which the characteristic energy scale $\Delta$ and the correlation length  $\xi$ are related by
\begin{equation}
\ln \Delta^{-1} \sim \xi^\psi, \qquad \psi = \frac{1}{2}.
\end{equation}
As a consequence, the standard KZ power law is expected to be replaced by logarithmic scaling, with the domain-wall density and residual energy density decaying as powers of $\ln \tau_{\rm ann}$~\cite{Dziarmaga2006,Caneva2007}. Moreover, the residual energy can decay faster than the domain-wall density because domain walls are not energetically equivalent in a disordered chain: defects on weak bonds contribute less energy than that on strong bonds.

For the quantum chain, this unconventional critical point admits a generalized Kibble--Zurek description of the dynamical  scaling~\cite{Dziarmaga2006}. The corresponding freeze-out length grows logarithmically with annealing time, giving for the domain wall density 
\begin{equation}
n_{\rm dw} \sim \left(\ln \tau_{\rm ann}\right)^{-2}.
\end{equation}
Our analysis, outlined in Appendix~\ref{app:disordered_chain}, shows that the residual energy density obeys a distinct but related scaling,
\begin{equation}
\epsilon_{\rm res} \sim n_{\rm dw}^2,
\label{eq:eres_vs_dw}
\end{equation}
and therefore
\begin{equation}
\epsilon_{\rm res} \sim \left(\ln \tau_{\rm ann}\right)^{-4}.
\end{equation}
For QA, these results follow from the established infinite-randomness theory. Our QIA numerical data are consistent with the same forms over the time window studied, although we are not aware of an analytical theory placing QIA in the same universality class. The origin of Eq.~\ref{eq:eres_vs_dw} lies in the statistics of low-energy defects: domain walls preferentially survive on weak bonds, so the residual energy decreases faster than the domain-wall density. 

\begin{figure}[t]
    \centering
    \includegraphics[width=0.8\linewidth]{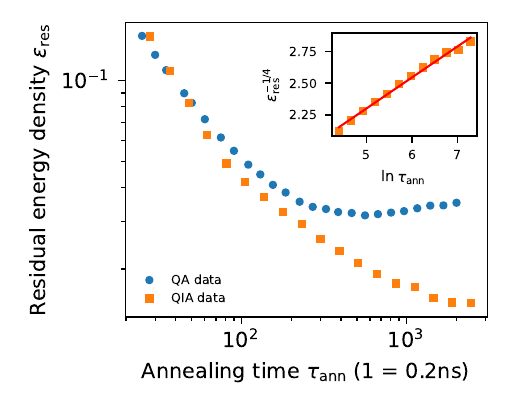}
    \caption{Residual energy density as a function of annealing time for the one-dimensional disordered transverse-field Ising chain with $N=1000$ and random ferromagnetic couplings $J_{i,i+1}\sim U(0,1)$. The QA results are obtained from the D-Wave annealer, while the QIA results are obtained from numerical integration of the QIA dynamics. In contrast to the clean chain, both protocols exhibit much slower relaxation, characteristic of infinite-randomness critical dynamics. The QIA data are consistent with logarithmic decay, as shown by the inset, where $\epsilon_{\rm res}^{-1/4}$ is plotted against $\ln \tau_{\rm ann}$. The QA data follow the same slow trend over the coherent-time regime accessible on the hardware, but flatten at longer annealing times due to decoherence and other open-system effects.}
    \label{fig:1d_random_tfim}
\end{figure}

\begin{figure}[t]
    \centering
    \includegraphics[width=0.8\linewidth]{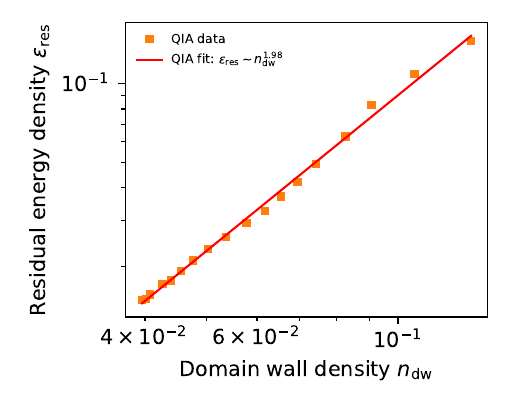}
    \caption{Residual energy density $\epsilon_{\mathrm{res}}$ as a function of domain-wall density $n_{\mathrm{dw}}$ for the one-dimensional disordered transverse-field Ising chain under QIA dynamics. The data show a clear power-law relation between the two quantities, with a fitted exponent $\epsilon_{\mathrm{res}} \sim n_{\mathrm{dw}}^{1.98}$, consistent with the expected quadratic scaling $\epsilon_{\mathrm{res}} \sim n_{\mathrm{dw}}^{2}$.}
    \label{fig:1d_random_eresvsndw}
\end{figure}

QIA simulations and quantum annealing experiments on a D-Wave annealer provide tests of these predictions. The residual energy density as a function of annealing time, shown in Fig.~\ref{fig:1d_random_tfim}, exhibits much slower relaxation than in the clean chain, consistent with the logarithmic behavior previously reported for quantum annealing of the random-bond  transverse-field Ising chain~\cite{Dziarmaga2006,Caneva2007}. In addition, the residual energy density plotted against the domain-wall density, shown in Fig.~\ref{fig:1d_random_eresvsndw}, confirms the quadratic relation $\epsilon_{\rm res}\sim n_{\rm dw}^2$ for the QIA dynamics. A similar quadratic relation between residual energy and defect density follows from classical cooling studies of random ferromagnetic Ising chains~\cite{Suzuki2009}. Within the coherent regime accessible on the D-Wave device, the QA data are also consistent with the same slow scaling behavior.

Taken together, these results show that strong disorder alone can suppress the role of quantum correlations in annealing dynamics, including in one-dimensional systems where the clean model exhibits a clear QA vs. QIA distinction. Despite the absence of frustration,
rare-region effects generate a common dynamical bottleneck for both protocols, reducing the distinction observed in the clean chain. This establishes disorder as a second key factor, independent of dimensionality, that controls whether quantum correlations  provide an advantageous dynamical scaling in QA.

\subsection{Two-dimensional spin glass}

The next comparison concerns the two-dimensional Edwards–Anderson spin-glass model~\cite{EdwardsAnderson1975} on a square lattice with open boundary conditions. The Hamiltonian is
\begin{equation}
H(t) = - A(t) \sum_i \sigma_i^x - B(t) \sum_{\langle i,j\rangle} J_{ij} \sigma_i^z \sigma_j^z ,
\end{equation}
where \(\langle i,j\rangle\) denotes nearest-neighbor pairs on the square lattice, and the couplings \(J_{ij}\) are independent random variables drawn from the uniform distribution \(J_{ij}\in [-1,1]\). In contrast to the one-dimensional disordered chain discussed above, the presence of both positive and negative couplings generates frustrated plaquettes and a complex energy landscape \footnote{Frustration in 1D can always be eliminated by a gauge transformation flipping all the spins say to the right of the frustrated bond}. This model therefore provides a natural setting for studying the combined effects of disorder and frustration on annealing dynamics.

Annealing performance is assessed by comparing the final configuration obtained from each protocol with the exact ground state of the corresponding final classical Ising Hamiltonian. For the two-dimensional planar Ising spin glass with no longitudinal fields, the exact ground state can be obtained efficiently by mapping the problem to a minimum-weight perfect matching problem and solving it with the blossom algorithm~\cite{Bieche1980,Barahona1982,Pardella2008}. Using this ground state as a reference, we measure the residual energy density \(\epsilon_{\rm res}\), defined in
Eq.~\ref{eq:eres}, together with a geometrical quantity that we call the residual circumference density.

To define this quantity, let \(s_i^{\rm f}\) be the final classical spin configuration obtained after annealing, and let \(s_i^0\) denote the ground-state configuration. Define the mismatch variable
\begin{equation}
q_i = s_i^{\rm f}s_i^0 ,
\end{equation}
so that \(q_i=+1\) if the final spin agrees with the ground state at site \(i\), and \(q_i=-1\) otherwise. The residual circumference density is then defined as
\begin{equation}
c_{\rm res}=
\frac{1}{N}
\sum_{\langle i,j\rangle}
\frac{1-q_i q_j}{2}.
\end{equation}
This quantity counts the total boundary length of the mismatch clusters per spin, namely the number of nearest-neighbor bonds separating regions where the final configuration and the ground state differ. While \(\epsilon_{\rm res}\) quantifies how far the final configuration lies above the optimum in energy, \(c_{\rm res}\) probes the spatial structure of the remaining defects.

\begin{figure}[t]
    \centering
    \includegraphics[width=0.8\linewidth]{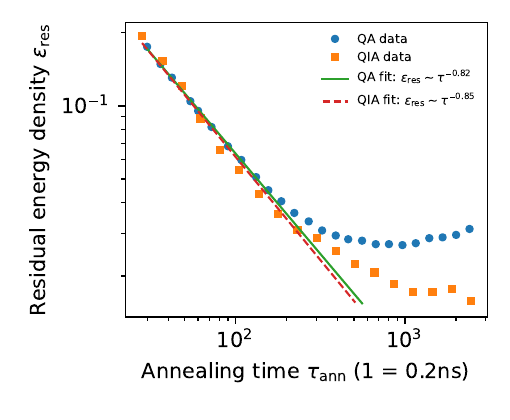}
    \caption{Residual energy density as a function of annealing time for a representative two-dimensional spin-glass instance on a $20\times 20$ square lattice. The QA data are obtained from the D-Wave annealer, while the QIA data are obtained numerically. In the intermediate-time regime, both protocols follow a common apparent power-law decay, with a fitted exponent close to $\epsilon_{\mathrm{res}} \sim \tau^{-0.82}$, indicating essentially identical scaling behavior within the coherent-time window accessible on the hardware. At longer annealing times, the QA data flatten and slightly increase, signaling the onset of decoherence and other open-system effects on the D-Wave device. By contrast, the QIA data continue to decrease, but much more slowly than the intermediate-time power law, consistent with a crossover toward glassy long-time relaxation.}
    \label{fig:2d_spinglass}
\end{figure}

\begin{figure}[t]
    \centering
    \includegraphics[width=0.8\linewidth]{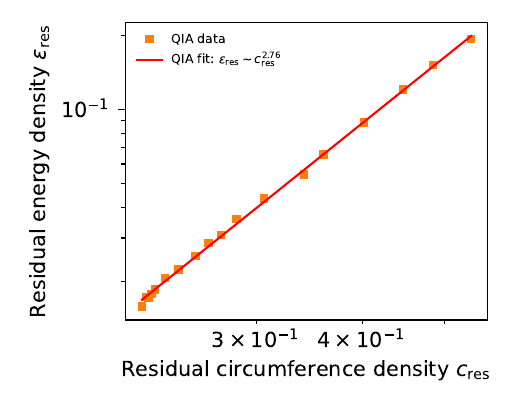}
    \caption{Residual energy density $\epsilon_{\mathrm{res}}$ as a function of residual circumference density $c_{\mathrm{res}}$ for the two-dimensional spin-glass system under QIA dynamics. The data follow a power-law relation, with fitted behavior $\epsilon_{\mathrm{res}} \sim c_{\mathrm{res}}^{2.76}$.}
    \label{fig:2d_spinglass_eresvsc}
\end{figure}

We compare QA implemented on the D-Wave annealer with numerical QIA simulation. For the \(20\times 20\) square lattice studied here, QA and QIA are performed on the same disorder realization, allowing a direct comparison between the two protocols, as shown in Fig.~\ref{fig:2d_spinglass}. In the intermediate annealing-time regime, both QA and QIA display an approximate power-law decay of the residual energy density, with essentially the same effective exponent for the two protocols. Previous simulated-annealing studies of two-dimensional Ising spin glasses reported smaller KZ scaling exponents~\cite{Rubin2017,Xu2017}. At longer annealing times, the two protocols exhibit somewhat different crossovers. The QA data from D-Wave eventually flatten and may even show a slight upturn, indicating that the dynamics have left the coherent regime and become limited by decoherence and other open-system effects on the hardware. By contrast, the QIA numerical data continue to decay, but much more slowly than in the intermediate-time power-law regime. This slower long-time behavior is consistent with a crossover toward glassy relaxation, possibly of the logarithmic form, as reported in earlier annealing studies of spin glass systems~\cite{Huse1986,santoro_2002}. Taken together, these results indicate that, in the 2D spin glass, QA does not exhibit a distinct scaling advantage over the quantum-correlation-free QIA baseline, suggesting that the combined effects of disorder and frustration limit the role of quantum correlations in the annealing dynamics. The data also exhibit an approximate power-law relation between the residual energy density and the residual circumference density, as shown in Fig.~\ref{fig:2d_spinglass_eresvsc}. This indicates a systematic correlation between the energy of the remaining excitations and the geometry of the mismatched domains.

\subsection{Three-dimensional spin glass}

The final case is a spin-glass instance defined on the Pegasus graph native to the D-Wave annealer~\cite{King2023,zhang2025computational}. The Pegasus graph provides a nonplanar, effectively three-dimensional connectivity structure and therefore represents a more complex setting than the two-dimensional planar spin glass studied above. In contrast to the planar case, exact ground-state computation for three-dimensional spin glasses is NP-hard, and no general efficient classical algorithm is available. The analysis is therefore restricted to specific problem instances for which the exact ground states were reported in previous studies~\cite{zhang2025computational}, allowing us to evaluate the annealing performance through the residual energy density relative to the ground state.

\begin{figure}[t]
    \centering
    \includegraphics[width=0.8\linewidth]{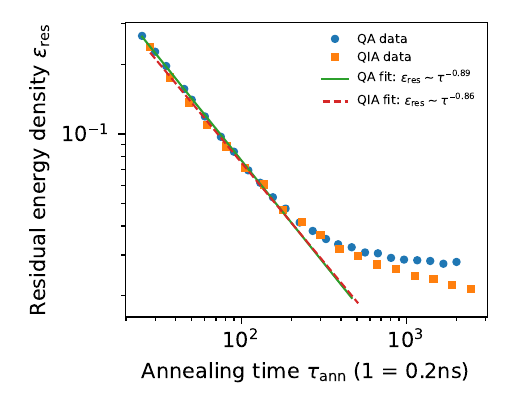}
    \caption{Residual energy density as a function of annealing time for a representative three-dimensional spin-glass instance defined on the Pegasus graph with $N=678$. The QA data are obtained from the D-Wave annealer, while the QIA data are obtained numerically. In the intermediate-time regime, both protocols exhibit nearly identical power-law decay, with a fitted exponent close to $0.9$, indicating no clear scaling advantage of QA over QIA within the coherent-time window accessible on the hardware. At longer annealing times, the QA data flatten because of decoherence and other open-system effects on the device, whereas the QIA data continue to decay slowly, consistent with a crossover toward glassy long-time relaxation.}
    \label{fig:3d_spinglass}
\end{figure}

Figure.~\ref{fig:3d_spinglass} compares QA on the D-Wave device with QIA for the same coupling, $J_{ij}$,  realization. Despite the greater complexity, the trends resemble those of the two-dimensional spin glass. In the coherent-time regime accessible on the hardware, QA and QIA exhibit nearly identical apparent power-law scaling of the residual energy. For the representative instance shown, both fitted exponents are close to $0.9$, consistent with the range reported in previous D-Wave studies of three-dimensional spin glasses~\cite{King2023}. Thus, the data show no clear scaling advantage of QA over the quantum-correlation-free baseline. At longer times, QA flattens as open-system effects become important, whereas the QIA residual energy continues to decrease more slowly, consistent with a crossover toward glassy long-time dynamics.

Together with the two-dimensional results, this suggests that in disordered and frustrated spin-glass systems, QA and QIA are limited by similar slow annealing dynamics. The three-dimensional spin-glass case therefore reinforces the broader conclusion that the quantum advantage is  constrained by dimensionality, disorder, and frustration.

\begin{table*}[t]
\caption{
Summary of the models and QA--QIA scaling comparisons. Clean 1D and LMG entries give theoretical leading powers. Clean 2D/3D and spin-glass entries describe empirical fits over accessible time windows.
}
\label{tab:model_summary}

\centering
\small
\renewcommand{\arraystretch}{1.25}
\setlength{\tabcolsep}{5pt}

\begin{tabular}{llccl}
\hline\hline

\parbox[t]{2.4cm}{\textbf{Model}}
&
\parbox[t]{2.7cm}{\textbf{Geometry or connectivity}}
&
\textbf{Disorder}
&
\textbf{Frustration}
&
\parbox[t]{6.2cm}{\textbf{Leading QA--QIA behavior}}
\\
\hline

\parbox[t]{2.4cm}{Clean 1D TFIM}
&
\parbox[t]{2.7cm}{One-dimensional, local}
&
No
&
No
&
\parbox[t]{6.2cm}{
Distinct algebraic scaling:
$\epsilon_{\rm res}^{\rm QA}\sim\tau_{\rm ann}^{-1/2}$ and
$\epsilon_{\rm res}^{\rm QIA}\sim\tau_{\rm ann}^{-1/3}$.
}
\\

\parbox[t]{2.4cm}{Clean 2D TFIM}
& \parbox[t]{2.7cm}{Square lattice, $30\times30$}
& No & No
& \parbox[t]{6.2cm}{Effective fitted exponents $\alpha_{\rm QA}\simeq0.74$ and $\alpha_{\rm QIA}\simeq0.64$; final-energy interpretation depends on post-critical evolution.} \\
\parbox[t]{2.4cm}{Clean 3D TFIM}
& \parbox[t]{2.7cm}{Cubic lattice, $10\times10\times10$}
& No & No
& \parbox[t]{6.2cm}{Effective fitted exponents $\alpha_{\rm QA}\simeq0.76$ and $\alpha_{\rm QIA}\simeq0.72$; smaller slope distinction than the 2D case.} \\

\parbox[t]{2.4cm}{LMG model}
&
\parbox[t]{2.7cm}{Fully connected, effective $d=\infty$}
&
No
&
No
&
\parbox[t]{6.2cm}{
Same leading algebraic scaling:
$\epsilon_{\rm res}^{\rm QA}\sim
\epsilon_{\rm res}^{\rm QIA}\sim\tau_{\rm ann}^{-1}$.
}
\\

\parbox[t]{2.4cm}{Random-bond 1D TFIM}
&
\parbox[t]{2.7cm}{One-dimensional, local}
&
Yes
&
No
&
\parbox[t]{6.2cm}{
QA follows the logarithmic scaling associated with infinite-randomness
criticality, while QIA displays numerically similar slow logarithmic
behavior.
}
\\

\parbox[t]{2.4cm}{2D Edwards--Anderson spin glass}
&
\parbox[t]{2.7cm}{Two-dimensional square lattice}
&
Yes
&
Yes
&
\parbox[t]{6.2cm}{
Same apparent intermediate-time scaling,
$\epsilon_{\rm res}\sim\tau_{\rm ann}^{-0.82}$, with no clear
QA--QIA scaling distinction.
}
\\

\parbox[t]{2.4cm}{Pegasus spin glass}
&
\parbox[t]{2.7cm}{Nonplanar, effectively three-dimensional}
&
Yes
&
Yes
&
\parbox[t]{6.2cm}{
Same apparent intermediate-time scaling,
$\epsilon_{\rm res}\sim\tau_{\rm ann}^{-0.9}$, with no clear
QA--QIA scaling distinction.
}
\\

\hline\hline
\end{tabular}
\end{table*}

\section{Discussion and Outlook}

Quantum-inspired annealing provides a classical baseline for assessing the role of quantum correlations in quantum annealing. QIA retains the Hamiltonian structure and annealing schedule while restricting the evolution to a product-state manifold. Our comparisons identify dimensionality, disorder, and frustration as key factors controlling the difference in residual-energy scaling between QA and QIA. The results are summarized in Table~\ref{tab:model_summary}. Here, a scaling advantage means a faster decay of residual energy with annealing time relative to QIA, rather than an end-to-end computational speedup over all classical algorithms.

The clean models show a diminishing distinction between QA and QIA with increasing dimensionality. In one dimension, their different critical theories yield the respective Kibble--Zurek scalings $\epsilon_{\rm res}^{\rm QA}\sim\tau_{\rm ann}^{-1/2}$ and
$\epsilon_{\rm res}^{\rm QIA}\sim\tau_{\rm ann}^{-1/3}$. The fitted exponents become closer in two and three dimensions, while both protocols share the leading scaling
$\epsilon_{\rm res}\sim\tau_{\rm ann}^{-1}$ in the fully connected LMG model. The higher-dimensional lattice results should nevertheless be distinguished from these theoretical predictions: final residual energy can depend on domain geometry and post-critical dynamics, including coarsening and excitation redistribution.

Disorder and frustration further reduce the observed scaling distinction. In the random-bond chain, weak bonds provide a common bottleneck, and both protocols exhibit similar logarithmic scaling over the annealing-time range studied. In the spin-glass examples, QA and QIA likewise display nearly identical power-law exponents over the fitted windows. Together, these results show that quantum correlations need not change the dominant scaling law, even when they are present in the quantum dynamics.

Several qualifications remain. Hardware QA is subject to environmental coupling, whereas QIA is non-dissipative, and the observed scaling need not persist beyond the fitted annealing-time windows. Moreover, agreement with QIA does not imply the absence of quantum correlations; it only shows that product-state dynamics can reproduce the scaling of the measured quantities. QIA therefore serves as a diagnostic baseline, not as the strongest classical optimizer. Comparisons with other classical and semiclassical methods remain important~\cite{Shin2014,Subires2022,Misra2023}.

Future work could systematically incorporate pair or cluster correlations into QIA to determine which correlations change the leading annealing behavior in clean models. Spatially inhomogeneous drivers, multispin interactions, and cyclic annealing~\cite{zhang2024cyclic} offer further ways to explore this question beyond the uniform transverse-field protocol studied here. Studies of system-size dependence, post-critical ordering, and environmental coupling would help establish the range of validity
of the observed scaling and clarify when quantum correlations provide an advantage.

\section{Acknowledgments}

We are grateful to Polina Matveeva and Hao Zhang for useful discussions. The work was supported by  the NSF grant DMR-2338819. Access to D-Wave quantum computing resources was supported through CloudBank allocation MTH260072. 

\appendix
\section{QIA Equations and Numerical Implementation}
\label{app:qia_equations}
Consider a Hamiltonian of a large spin-$S$ model
\begin{equation}
H
=
-A\sum_i s_i^x
-B\sum_{(i,j)\in E}J_{ij}s_i^zs_j^z,
\end{equation}
where
\begin{equation}
s_i^\alpha=\frac{S_i^\alpha}{S}.
\end{equation}
Setting $\hbar=1$, the normalized spin operators satisfy the commutation relations 
\begin{equation}
[s_i^\alpha,s_j^\beta]
=
\frac{i}{S}
\delta_{ij}\epsilon_{\alpha\beta\gamma}s_i^\gamma.
\end{equation}
Using the Heisenberg equation,
\begin{equation}
\frac{d s_i^\alpha}{dt}
=
i[H,s_i^\alpha],
\end{equation}
the transverse-field contribution is
\begin{align}
i\left[-A\sum_k s_k^x,s_i^\alpha\right]
&=
-iA[s_i^x,s_i^\alpha]
\nonumber\\
&=
\frac{A}{S}
\epsilon_{\alpha\beta x}s_i^\beta.
\end{align}
Similarly, the interaction contribution is
\begin{align}
&i\left[
-B\sum_{(k,l)\in E}J_{kl}s_k^zs_l^z,
s_i^\alpha
\right]
\nonumber\\
&\quad
=
-iB\sum_{(k,l)\in E}J_{kl}
\left(
s_k^z[s_l^z,s_i^\alpha]
+
[s_k^z,s_i^\alpha]s_l^z
\right)
\nonumber\\
&\quad
=
\frac{B}{S}
\epsilon_{z\alpha\beta}
\sum_{(k,l)\in E}J_{kl}
\left(
\delta_{li}s_k^zs_i^\beta
+
\delta_{ki}s_i^\beta s_l^z
\right)
\nonumber\\
&\quad
=
\frac{B}{S}
\epsilon_{\alpha\beta z}s_i^\beta
\sum_{j\in\partial i}J_{ij}s_j^z.
\end{align}
Combining the two contributions gives
\begin{equation}
\frac{d s_i^\alpha}{dt}
=
\frac{A}{S}
\epsilon_{\alpha\beta x}s_i^\beta
+
\frac{B}{S}
\epsilon_{\alpha\beta z}s_i^\beta
\sum_{j\in\partial i}J_{ij}s_j^z.
\end{equation}
In components, these equations read
\begin{equation}
\begin{aligned}
\frac{d s_i^x}{dt}
&=
\frac{B}{S}s_i^y
\sum_{j\in\partial i}J_{ij}s_j^z,
\\
\frac{d s_i^y}{dt}
&=
\frac{A}{S}s_i^z
-
\frac{B}{S}s_i^x
\sum_{j\in\partial i}J_{ij}s_j^z,
\\
\frac{d s_i^z}{dt}
&=
-\frac{A}{S}s_i^y.
\end{aligned}
\end{equation}
Applying the product-state approximation,
$\langle s_i^{\alpha}s_j^{\beta}\rangle = \langle s_i^{\alpha} \rangle \langle s_j^{\beta} \rangle=m_i^{\alpha}m_j^{\beta}$,
and setting $S=\frac{1}{2}$ recovers Eq.~\eqref{eq:qia_eom}. To seed the instability associated with quantum fluctuations, we use the initial conditions
\begin{equation}
    \vec{m}_i(t=0) = \left(\cos\theta_i,\,\sin\theta_i\sin\phi_i,\,\sin\theta_i\cos\phi_i\right),
\end{equation}
where $\theta_i$ and $\phi_i$ are independent random variables drawn from
\begin{equation}
\theta_i \sim \operatorname{Uniform}(0,\theta_{\max}),
\qquad
\phi_i \sim \operatorname{Uniform}(0,2\pi),
\end{equation}
where $\theta_{max} = 0.1$ throughout this work.

\section{Critical scaling of QIA in clean transverse-field Ising models}
\label{app:qia_1d_critical}

In this Appendix, we first derive the critical exponents governing QIA
in the clean one-dimensional transverse-field Ising model, and then
extend the same discussion to square and cubic lattices.
Following the spin-$S$ formulation introduced in Appendix \ref{app:qia_equations}, we consider
\begin{equation}
H
=
-A\sum_i \frac{S_i^x}{S}
-B\sum_i \frac{S_i^z S_{i+1}^z}{S^2},
\label{eq:app_clean_spinS}
\end{equation}
where $S_i^\alpha$ are spin-$S$ operators. Throughout this Appendix
we take the uniform coupling $J=1$, as in Sec.~III\,A.

On the transverse-field-dominated side of the transition, the
classical spins are polarized along the positive $x$ direction.
We therefore perform a Holstein--Primakoff expansion using the
$x$ axis as the quantization axis,
\begin{align}
S_i^x &= S-a_i^\dagger a_i,
\\
S_i^+ &= \sqrt{2S-a_i^\dagger a_i}\,a_i,
\\
S_i^- &= a_i^\dagger\sqrt{2S-a_i^\dagger a_i},
\end{align}
where the raising and lowering operators are defined with respect to
the $x$ axis. For $S\gg 1$, keeping only the leading terms gives
\begin{equation}
S_i^+
\simeq
\sqrt{2S}\,a_i,
\qquad
S_i^-
\simeq
\sqrt{2S}\,a_i^\dagger.
\end{equation}
Using
\begin{equation}
S_i^z=\frac{S_i^+ + S_i^-}{2},
\end{equation}
we obtain
\begin{equation}
S_i^z
\simeq
\sqrt{\frac{S}{2}}
\left(a_i+a_i^\dagger\right).
\end{equation}

Substituting these expressions into Eq.~\eqref{eq:app_clean_spinS}
and retaining terms up to quadratic order in the bosonic operators
gives
\begin{equation}
H
=
\frac{A}{S}\sum_i a_i^\dagger a_i
-
\frac{B}{2S}
\sum_i
\left(a_i+a_i^\dagger\right)
\left(a_{i+1}+a_{i+1}^\dagger\right)
+\mathrm{const.}
\label{eq:app_hp_quadratic}
\end{equation}
The constant term will be dropped below. The overall factor $1/S$
sets the time or energy scale of the semiclassical dynamics and does
not affect the critical exponents.

To diagonalize Eq.~\eqref{eq:app_hp_quadratic}, we introduce the
Fourier transformation
\begin{equation}
a_j
=
\frac{1}{\sqrt{N}}
\sum_k a_k e^{ikj},
\qquad
a_j^\dagger
=
\frac{1}{\sqrt{N}}
\sum_k a_k^\dagger e^{-ikj}.
\label{eq:app_fourier}
\end{equation}
In momentum space, the quadratic Hamiltonian can be written as
\begin{equation}
H
=
\sum_k
\left[
\alpha_k a_k^\dagger a_k
+
\frac{1}{2}
\left(
\beta_k a_k a_{-k}
+
\beta_k^* a_k^\dagger a_{-k}^\dagger
\right)
\right],
\label{eq:app_momentum_hamiltonian}
\end{equation}
where
\begin{equation}
\alpha_k
=
\frac{A-B\cos k}{S}.
\end{equation}
After combining the $k$ and $-k$ contributions, the anomalous
coefficient can be chosen real,
\begin{equation}
\widetilde{\beta}_k
=
-\frac{B}{S}\cos k.
\end{equation}

The Hamiltonian is diagonalized by a bosonic Bogoliubov
transformation,
\begin{equation}
b_k
=
u_k a_k
+
v_k a_{-k}^\dagger,
\qquad
|u_k|^2-|v_k|^2=1.
\end{equation}
The resulting excitation spectrum is
\begin{align}
E_k
&=
\sqrt{
\alpha_k^2-\widetilde{\beta}_k^2
}
\nonumber\\
&=
\frac{1}{S}
\sqrt{
\left(A-B\cos k\right)^2
-
B^2\cos^2 k
}
\nonumber\\
&=
\frac{1}{S}
\sqrt{
A\left(A-2B\cos k\right)
}.
\label{eq:app_qia_spectrum}
\end{align}

The minimum of the spectrum occurs at $k=0$, so the excitation gap is
\begin{equation}
\Delta
=
E_{k=0}
=
\frac{1}{S}
\sqrt{A(A-2B)}.
\label{eq:app_qia_gap}
\end{equation}
The gap therefore closes at
\begin{equation}
A_c=2B.
\label{eq:app_qia_criticalpoint}
\end{equation}
For the linear annealing schedule used in the main text,
\begin{equation}
A=1-f,
\qquad
B=f,
\end{equation}
this corresponds to
\begin{equation}
f_c=\frac{1}{3}.
\end{equation}

To extract the critical exponents, we define the distance from the
critical point as
\begin{equation}
\delta=A-2B.
\end{equation}
Close to the transition, $A$ remains finite and
Eq.~\eqref{eq:app_qia_gap} gives
\begin{equation}
\Delta
\sim
|\delta|^{1/2}.
\label{eq:app_gap_scaling}
\end{equation}
Using the general critical scaling relation
\begin{equation}
\Delta
\sim
\xi^{-z}
\sim
|\delta|^{z\nu},
\end{equation}
we therefore obtain
\begin{equation}
z\nu=\frac{1}{2}.
\label{eq:app_znu}
\end{equation}

The dynamical exponent can be determined independently from the
dispersion at criticality. Setting $A=2B$ in
Eq.~\eqref{eq:app_qia_spectrum} gives
\begin{equation}
E_k
=
\frac{2B}{S}
\sqrt{1-\cos k}.
\end{equation}
For $k\ll 1$,
\begin{equation}
1-\cos k
\simeq
\frac{k^2}{2},
\end{equation}
and hence
\begin{equation}
E_k\propto |k|.
\end{equation}
Comparing with the critical dispersion
\begin{equation}
E_k\sim |k|^z,
\end{equation}
we find
\begin{equation}
z_{\rm QIA}=1.
\label{eq:app_qia_z}
\end{equation}
Together with Eq.~\eqref{eq:app_znu}, this yields
\begin{equation}
\nu_{\rm QIA}=\frac{1}{2}.
\label{eq:app_qia_nu}
\end{equation}

The QIA dynamics of the clean one-dimensional chain is therefore
characterized by the semiclassical critical exponents
\begin{equation}
\boxed{
\nu_{\rm QIA}=\frac{1}{2},
\qquad
z_{\rm QIA}=1.
}
\label{eq:app_qia_exponents}
\end{equation}

These exponents determine the defect-production scaling through the
Kibble--Zurek mechanism. For a linear sweep through the critical
point, the frozen correlation length scales as
\begin{equation}
\hat{\xi}
\sim
\tau_{\rm ann}^{\frac{\nu}{1+z\nu}}.
\end{equation}
Using Eq.~\eqref{eq:app_qia_exponents}, we obtain
\begin{equation}
\hat{\xi}_{\rm QIA}
\sim
\tau_{\rm ann}^{1/3}.
\end{equation}
In one dimension, the domain-wall density is inversely proportional
to the characteristic domain size,
\begin{equation}
n_{\rm dw}^{\rm QIA}
\sim
\hat{\xi}_{\rm QIA}^{-1}
\sim
\tau_{\rm ann}^{-1/3}.
\label{eq:app_qia_ndw}
\end{equation}
For the clean ferromagnetic chain, each domain wall carries the same
energy cost, so the residual energy density is proportional to the
domain-wall density. Therefore,
\begin{equation}
\boxed{
\epsilon_{\rm res}^{\rm QIA}
\sim
n_{\rm dw}^{\rm QIA}
\sim
\tau_{\rm ann}^{-1/3}.
}
\label{eq:app_qia_residual}
\end{equation}
This gives the QIA scaling used in Sec.~III\,A.

\subsection{Extension to square and cubic lattices}
\label{app:qia_higher_d}

The same large-\(S\) expansion applies to a \(d\)-dimensional
hypercubic lattice. With the lattice spacing and \(J\) set to
unity, the structure factor
\begin{equation}
\gamma_{\mathbf k}=\sum_{\mu=1}^{d}\cos k_\mu
\end{equation}
replaces \(\cos k\) in the one-dimensional calculation.
The quadratic coefficients become
\(\alpha_{\mathbf k}=(A-B\gamma_{\mathbf k})/S\) and
\(\widetilde{\beta}_{\mathbf k}=-B\gamma_{\mathbf k}/S\),
yielding
\begin{equation}
E_{\mathbf k}
=\frac{1}{S}\sqrt{A\left(A-2B\gamma_{\mathbf k}\right)}.
\label{eq:app_qia_bulk_dispersion}
\end{equation}

The bulk gap closes at \(A_c=2dB_c\), corresponding to
\(f_c=1/(1+2d)\) for the linear schedule.
Writing \(\delta=A-2dB>0\) and expanding at small momentum gives
\begin{equation}
E_{\mathbf k}^{\,2}
=\frac{A}{S^2}
\left[\delta+B|\mathbf k|^2+O(|\mathbf k|^4)\right].
\end{equation}
Thus, \(\Delta\sim\delta^{1/2}\), the critical dispersion
satisfies \(E_{\mathbf k}\sim|\mathbf k|\), and balancing
\(\delta\) with \(B|\mathbf k|^2\) gives
\(\xi\sim\delta^{-1/2}\).
The resulting exponents are
\(\nu_{\mathrm{QIA}}=1/2\) and \(z_{\mathrm{QIA}}=1\)
for both square and cubic lattices, giving the KZ estimate
\(\hat\xi_{\mathrm{QIA}}\sim\tau_{\mathrm{ann}}^{1/3}\).
This calculation describes the translation-invariant bulk;
boundary corrections for the finite open samples are not included.

\section{Critical dynamics of the Lipkin--Meshkov--Glick model}
\label{app:lmg}

In this Appendix, we derive the asymptotic scaling of the residual
energy for QIA in the LMG model. The Hamiltonian considered in
Sec.~III\,B is
\begin{equation}
H
=
-A(t)\sum_i \sigma_i^x
-
\frac{B(t)}{N}
\sum_{i<j}\sigma_i^z\sigma_j^z.
\label{eq:app_lmg_hamiltonian}
\end{equation}
Introducing the collective spin
\begin{equation}
S^\alpha=\frac{1}{2}\sum_i\sigma_i^\alpha,
\qquad
S=\frac{N}{2},
\end{equation}
and using
\begin{equation}
\sum_{i<j}\sigma_i^z\sigma_j^z
=
2(S^z)^2-\frac{N}{2},
\end{equation}
Eq.~\eqref{eq:app_lmg_hamiltonian} can be written, up to an
irrelevant constant, as
\begin{equation}
H
=
-2A S^x
-
\frac{B}{S}(S^z)^2.
\label{eq:app_lmg_collective}
\end{equation}

In the classical limit, we introduce the normalized collective spin
$\mathbf{n}=(x,y,z)=\mathbf{S}/S$, with
$x^2+y^2+z^2=1$. The energy per spin is
\begin{equation}
e
=
-Ax-\frac{B}{2}z^2.
\label{eq:app_lmg_energy}
\end{equation}
A convenient parametrization is
\begin{equation}
x=\sqrt{1-z^2}\cos\phi,
\qquad
y=\sqrt{1-z^2}\sin\phi.
\end{equation}
Using $\mathcal{H}=2e$ as the canonical Hamiltonian for the pair
$(\phi,z)$ gives
\begin{equation}
\mathcal{H}
=
-2A\sqrt{1-z^2}\cos\phi-Bz^2,
\label{eq:app_lmg_canonical}
\end{equation}
which reproduces the collective QIA equations of motion.

Close to the paramagnetic fixed point, $z,\phi\ll1$,
Eq.~\eqref{eq:app_lmg_canonical} can be expanded as
\begin{equation}
\mathcal{H}
=
A\phi^2
+
(A-B)z^2
+
\frac{A}{4}z^4
+\cdots,
\label{eq:app_lmg_quartic}
\end{equation}
where an additive constant and higher-order terms have been omitted.
The quadratic coefficient changes sign at
\begin{equation}
A_c=B_c.
\end{equation}
For the linear schedule $A=1-f$, $B=f$, this corresponds to
$f_c=1/2$.

We now measure time from the critical point,
$t\rightarrow t-t_c$, with $t_c=\tau_{\rm ann}/2$. Near the
transition,
\begin{equation}
A-B=-\frac{2t}{\tau_{\rm ann}},
\qquad
A\simeq A_c=\frac{1}{2},
\end{equation}
so that Eq.~\eqref{eq:app_lmg_quartic} takes the universal quartic
form
\begin{equation}
\mathcal{H}_{\rm eff}
\simeq
\frac{1}{2}\phi^2
-
\frac{2t}{\tau_{\rm ann}}z^2
+
\frac{1}{8}z^4.
\label{eq:app_lmg_quartic_effective}
\end{equation}
The corresponding equation of motion is the normal form of a
nondissipative passage through a quartic critical point and can be
reduced, after rescaling, to the Painlev\'e-II equation.

To determine the characteristic scales, we write
\begin{equation}
t\sim\tau_{\rm ann}^{\alpha},
\qquad
z\sim\tau_{\rm ann}^{-\beta}.
\end{equation}
Balancing the inertial, time-dependent quadratic, and nonlinear terms
in the equation of motion gives
\begin{equation}
\alpha=\beta=\frac{1}{3}.
\end{equation}
Therefore,
\begin{equation}
t\sim\tau_{\rm ann}^{1/3},
\qquad
z\sim\tau_{\rm ann}^{-1/3}.
\label{eq:app_lmg_scaling_z}
\end{equation}
Since the conjugate momentum satisfies $\phi\sim\dot z$, it scales as
\begin{equation}
\phi\sim\tau_{\rm ann}^{-2/3}.
\label{eq:app_lmg_scaling_phi}
\end{equation}

The nonadiabatic excitation generated while crossing the critical
region is conveniently characterized by the classical adiabatic
invariant
\begin{equation}
I
=
\frac{1}{2\pi}\left|\oint \phi\,dz \right|.
\label{eq:app_lmg_action}
\end{equation}
Using Eqs.~\eqref{eq:app_lmg_scaling_z} and
\eqref{eq:app_lmg_scaling_phi}, its characteristic scale is
\begin{equation}
I_{\rm exc}
\sim
\tau_{\rm ann}^{-2/3}
\tau_{\rm ann}^{-1/3}
=
\tau_{\rm ann}^{-1}.
\label{eq:app_lmg_action_scaling}
\end{equation}
Equivalently, the Painlev\'e rescaling removes the explicit annealing
time from the dimensionless critical dynamics, while the physical
phase-space area acquires an overall factor
$\tau_{\rm ann}^{-1}$.

After the system leaves the critical region, the motion becomes
approximately harmonic around one of the symmetry-broken minima.
The excitation energy is then proportional to the adiabatic
invariant,
\begin{equation}
E_{\rm exc}\simeq\omega I_{\rm exc},
\end{equation}
where the final oscillation frequency is finite and independent of
$\tau_{\rm ann}$. Consequently, the residual energy density obeys
the leading algebraic scaling
\begin{equation}
\boxed{
\epsilon_{\rm res}^{\rm QIA}
\sim
\tau_{\rm ann}^{-1}.
}
\label{eq:app_lmg_residual_scaling}
\end{equation}
Weak dependence on the initial fluctuation amplitude or logarithmic
corrections does not modify this leading power-law exponent.

\section{Scaling of defects and residual energy in the
one-dimensional disordered chain}
\label{app:disordered_chain}

In this Appendix, we summarize the scaling argument for the
one-dimensional disordered transverse-field Ising chain considered
in Sec.~III\,C. We first obtain the logarithmic dependence of the
domain-wall density from the activated critical dynamics, and then
show that the statistics of weak bonds lead to the relation
$\epsilon_{\rm res}\sim n_{\rm dw}^2$.

For the random transverse-field Ising chain, the infinite-randomness
critical point is characterized by
\begin{equation}
\xi\sim |\delta|^{-\nu},
\qquad
\ln \tau_{\rm rel}\sim \xi^\psi,
\label{eq:app_random_activated}
\end{equation}
where $\delta$ denotes the distance from the critical point. For the
one-dimensional random Ising universality class~\cite{Fisher1995,Igloi2001,Igloi2002},
\begin{equation}
\nu=2,
\qquad
\psi=\frac{1}{2}.
\end{equation}
Combining these relations gives
\begin{equation}
\ln \tau_{\rm rel}
\sim
|\delta|^{-\nu\psi}
=
|\delta|^{-1}.
\label{eq:app_random_relaxation}
\end{equation}

For a linear annealing protocol,
$\delta\propto (t-t_c)/\tau_{\rm ann}$, the time scale associated
with the variation of the control parameter is
\begin{equation}
\tau_{\rm drive}
\sim
\left|
\frac{\delta}{\dot{\delta}}
\right|
\sim
\tau_{\rm ann}|\delta|.
\end{equation}
The freeze-out point $\hat{\delta}$ is determined by
$\tau_{\rm rel}\sim\tau_{\rm drive}$. Using
Eq.~\eqref{eq:app_random_relaxation}, this gives asymptotically
\begin{equation}
\frac{c}{|\hat{\delta}|}
\sim
\ln\left(
\tau_{\rm ann}|\hat{\delta}|
\right),
\end{equation}
where $c$ is a nonuniversal constant. To leading logarithmic order,
\begin{equation}
|\hat{\delta}|
\sim
\frac{1}{\ln\tau_{\rm ann}}.
\label{eq:app_random_delta}
\end{equation}
The corresponding frozen correlation length therefore scales as
\begin{equation}
\hat{\xi}
\sim
|\hat{\delta}|^{-\nu}
\sim
(\ln\tau_{\rm ann})^2.
\label{eq:app_random_xi}
\end{equation}
Since the typical distance between surviving domain walls is of
order $\hat{\xi}$, their density is
\begin{equation}
\boxed{
n_{\rm dw}
\sim
\hat{\xi}^{-1}
\sim
(\ln\tau_{\rm ann})^{-2}.
}
\label{eq:app_random_ndw}
\end{equation}

Unlike the clean chain, domain walls in the disordered system do not
carry a fixed energy. Since the open one-dimensional chain contains
no frustration, the signs of the couplings can be gauged away and
we may take $J_i>0$. A low-energy domain wall preferentially resides
on a locally weak bond. To characterize this effect, consider a
bond that is the weakest among $2k+1$ neighboring bonds. For
independent couplings drawn from a continuous distribution with
probability density $p(J)$ and cumulative distribution $F(J)$, the
minimum
\begin{equation}
J_{\min}
=
\min\{J_1,\ldots,J_{2k+1}\}
\end{equation}
has survival probability
\begin{equation}
{\rm Prob}(J_{\min}>J)
=
[1-F(J)]^{2k+1}.
\end{equation}
Its expectation value can therefore be written as
\begin{equation}
\langle J_{\min}\rangle
=
\int_0^\infty
[1-F(J)]^{2k+1}\,dJ.
\label{eq:app_random_minimum}
\end{equation}
For large $k$, the integral is controlled by the small-$J$ behavior.
If $p(0^+)$ is finite and nonzero,
\begin{equation}
F(J)\simeq p(0^+)J,
\end{equation}
and Eq.~\eqref{eq:app_random_minimum} gives
\begin{equation}
\langle J_{\min}\rangle
\sim
\frac{1}{k}.
\label{eq:app_random_weakbond}
\end{equation}
For the uniform distribution $J_i\sim U(0,1)$ used in this work,
$p(0^+)=1$.

After annealing, the relevant length scale is the frozen correlation
length $\hat{\xi}$. A surviving domain wall is therefore expected
to be localized on a weak bond selected over a region of size
$k\sim\hat{\xi}$. Its characteristic energy consequently scales as
\begin{equation}
\epsilon_{\rm dw}
\sim
\frac{1}{\hat{\xi}}.
\label{eq:app_random_dwenergy}
\end{equation}
Since the residual energy density is the product of the density of
domain walls and their characteristic energy,
\begin{equation}
\epsilon_{\rm res}
\sim
n_{\rm dw}\epsilon_{\rm dw}
\sim
\hat{\xi}^{-2}.
\end{equation}
Using Eq.~\eqref{eq:app_random_xi}, we obtain
\begin{equation}
\boxed{
\epsilon_{\rm res}
\sim
(\ln\tau_{\rm ann})^{-4}.
}
\label{eq:app_random_eres}
\end{equation}
Equivalently, Eqs.~\eqref{eq:app_random_ndw} and
\eqref{eq:app_random_dwenergy} imply the direct relation
\begin{equation}
\boxed{
\epsilon_{\rm res}
\sim
n_{\rm dw}^2,
}
\label{eq:app_random_energy_domainwall}
\end{equation}
which is independent of the detailed dependence of the frozen
length on the annealing time.

\bibliography{refs}

@article{kadowaki_1998,
  title = {Quantum annealing in the transverse Ising model},
  author = {Kadowaki, Tadashi and Nishimori, Hidetoshi},
  journal = {Phys. Rev. E},
  volume = {58},
  issue = {5},
  pages = {5355--5363},
  numpages = {0},
  year = {1998},
  month = {Nov},
  publisher = {American Physical Society},
  doi = {10.1103/PhysRevE.58.5355},
  url = {https://link.aps.org/doi/10.1103/PhysRevE.58.5355}
}

@article{brooke_1999,
author = {J. Brooke  and D. Bitko  and T. F. Rosenbaum and G. Aeppli },
title = {Quantum Annealing of a Disordered Magnet},
journal = {Science},
volume = {284},
number = {5415},
pages = {779-781},
year = {1999},
doi = {10.1126/science.284.5415.779},
URL = {https://www.science.org/doi/abs/10.1126/science.284.5415.779}
}

@article{santoro_2002,
author = {Giuseppe E. Santoro  and Roman Martoňák  and Erio Tosatti  and Roberto Car },
title = {Theory of Quantum Annealing of an Ising Spin Glass},
journal = {Science},
volume = {295},
number = {5564},
pages = {2427-2430},
year = {2002},
doi = {10.1126/science.1068774},
URL = {https://www.science.org/doi/abs/10.1126/science.1068774}}

@article{farhi_2000,
  title={Quantum computation by adiabatic evolution},
  author={Farhi, Edward and Goldstone, Jeffrey and Gutmann, Sam and Sipser, Michael},
  journal={arXiv preprint quant-ph/0001106},
  year={2000}
}

@article{morita_2008,
  title={Mathematical foundation of quantum annealing},
  author={Morita, Satoshi and Nishimori, Hidetoshi},
  journal={Journal of Mathematical Physics},
  volume={49},
  number={12},
  year={2008},
  publisher={AIP Publishing}
}

@article{Albash_2018,
  title = {Adiabatic quantum computation},
  author = {Albash, Tameem and Lidar, Daniel A.},
  journal = {Rev. Mod. Phys.},
  volume = {90},
  issue = {1},
  pages = {015002},
  numpages = {64},
  year = {2018},
  month = {Jan},
  publisher = {American Physical Society},
  doi = {10.1103/RevModPhys.90.015002},
  url = {https://link.aps.org/doi/10.1103/RevModPhys.90.015002}
}

@article{Ray1989,
  title = {Sherrington-Kirkpatrick model in a transverse field: Absence of replica symmetry breaking due to quantum fluctuations},
  author = {Ray, P. and Chakrabarti, B. K. and Chakrabarti, A.},
  journal = {Physical Review B},
  volume = {39},
  number = {16},
  pages = {11828--11832},
  year = {1989},
  publisher = {American Physical Society},
  doi = {10.1103/PhysRevB.39.11828}
}

@article{Das2008,
  title = {Colloquium: Quantum annealing and analog quantum computation},
  author = {Das, Arnab and Chakrabarti, Bikas K.},
  journal = {Rev. Mod. Phys.},
  volume = {80},
  issue = {3},
  pages = {1061--1081},
  numpages = {0},
  year = {2008},
  month = {Sep},
  publisher = {American Physical Society},
  doi = {10.1103/RevModPhys.80.1061},
  url = {https://link.aps.org/doi/10.1103/RevModPhys.80.1061}
}

@article{Katzgraber2014,
  title = {Glassy Chimeras Could Be Blind to Quantum Speedup: Designing Better Benchmarks for Quantum Annealing Machines},
  author = {Katzgraber, Helmut G. and Hamze, Firas and Andrist, Ruben S.},
  journal = {Physical Review X},
  volume = {4},
  number = {2},
  pages = {021008},
  year = {2014},
  publisher = {American Physical Society},
  doi = {10.1103/PhysRevX.4.021008}
}

@article{Heim2015,
  title = {Quantum versus Classical Annealing of Ising Spin Glasses},
  author = {Heim, Bettina and R{\o}nnow, Troels F. and Isakov, Sergei V. and Troyer, Matthias},
  journal = {Science},
  volume = {348},
  number = {6231},
  pages = {215--217},
  year = {2015},
  publisher = {American Association for the Advancement of Science},
  doi = {10.1126/science.aaa4170}
}

@article{Denchev2016,
  title = {What is the Computational Value of Finite-Range Tunneling?},
  author = {Denchev, Vasil S. and Boixo, Sergio and Isakov, Sergei V. and Ding, Nan and Babbush, Ryan and Smelyanskiy, Vadim and Martinis, John and Neven, Hartmut},
  journal = {Physical Review X},
  volume = {6},
  number = {3},
  pages = {031015},
  year = {2016},
  publisher = {American Physical Society},
  doi = {10.1103/PhysRevX.6.031015}
}

@article{Mandra2016,
  title = {Strengths and weaknesses of weak-strong cluster problems: A detailed overview of state-of-the-art classical heuristics versus quantum approaches},
  author = {Mandr{\`a}, Salvatore and Zhu, Zheng and Wang, Wenlong and Perdomo-Ortiz, Alejandro and Katzgraber, Helmut G.},
  journal = {Physical Review A},
  volume = {94},
  number = {2},
  pages = {022337},
  year = {2016},
  publisher = {American Physical Society},
  doi = {10.1103/PhysRevA.94.022337}
}

@inproceedings{Crosson2016,
  title = {Simulated Quantum Annealing Can Be Exponentially Faster than Classical Simulated Annealing},
  author = {Crosson, Elizabeth and Harrow, Aram W.},
  booktitle = {2016 IEEE 57th Annual Symposium on Foundations of Computer Science (FOCS)},
  pages = {714--723},
  year = {2016},
  organization = {IEEE},
  eprint = {1601.03030},
  archivePrefix = {arXiv},
  primaryClass = {quant-ph}
}

@article{Johnson2011,
  title = {Quantum annealing with manufactured spins},
  author = {Johnson, M. W. and Amin, M. H. S. and Gildert, S. and Lanting, T. and Hamze, F. and Dickson, N. and Harris, R. and Berkley, A. J. and Johansson, J. and Bunyk, P. and others},
  journal = {Nature},
  volume = {473},
  number = {7346},
  pages = {194--198},
  year = {2011},
  doi = {10.1038/nature10012}
}

@article{Dickson2013,
  title = {Thermally assisted quantum annealing of a 16-qubit problem},
  author = {Dickson, N. G. and Johnson, M. W. and Amin, M. H. S. and Harris, R. and Altomare, F. and Berkley, A. J. and Bunyk, P. and Cai, J. and Chapple, E. M. and Chavez, P. and others},
  journal = {Nature Communications},
  volume = {4},
  pages = {1903},
  year = {2013},
  doi = {10.1038/ncomms2920}
}

@article{Boixo2014,
  title = {Evidence for quantum annealing with more than one hundred qubits},
  author = {Boixo, Sergio and R{\o}nnow, Troels F. and Isakov, Sergei V. and Wang, Zhihui and Wecker, David and Lidar, Daniel A. and Martinis, John M. and Troyer, Matthias},
  journal = {Nature Physics},
  volume = {10},
  pages = {218--224},
  year = {2014},
  doi = {10.1038/nphys2900}
}

@article{Lanting2014,
  title = {Entanglement in a Quantum Annealing Processor},
  author = {Lanting, T. and Przybysz, A. J. and Smirnov, A. Yu. and Spedalieri, F. M. and Amin, M. H. and Berkley, A. J. and Harris, R. and Altomare, F. and Boixo, S. and Bunyk, P. and others},
  journal = {Physical Review X},
  volume = {4},
  number = {2},
  pages = {021041},
  year = {2014},
  doi = {10.1103/PhysRevX.4.021041}
}

@article{Ronnow2014,
  title = {Defining and detecting quantum speedup},
  author = {R{\o}nnow, Troels F. and Wang, Zhihui and Job, Joshua and Boixo, Sergio and Isakov, Sergei V. and Wecker, David and Martinis, John M. and Lidar, Daniel A. and Troyer, Matthias},
  journal = {Science},
  volume = {345},
  number = {6195},
  pages = {420--424},
  year = {2014},
  doi = {10.1126/science.1252319}
}

@article{Harris2018,
  title = {Phase transitions in a programmable quantum spin glass simulator},
  author = {Harris, R. and Sato, Y. and Berkley, A. J. and Reis, M. and Altomare, F. and Amin, M. H. and Boothby, K. and Bunyk, P. and Deng, C. and Enderud, C. and others},
  journal = {Science},
  volume = {361},
  number = {6398},
  pages = {162--165},
  year = {2018},
  doi = {10.1126/science.aat2025}
}

@article{Bando2020,
  title = {Probing the universality of topological defect formation in a quantum annealer: Kibble-Zurek mechanism and beyond},
  author = {Bando, Yuki and Susa, Yuki and Oshiyama, Hiroki and Shibata, Naokazu and Ohzeki, Masayuki and G\'omez-Ruiz, Fernando Javier and Lidar, Daniel A. and Suzuki, Sei and del Campo, Adolfo and Nishimori, Hidetoshi},
  journal = {Phys. Rev. Res.},
  volume = {2},
  issue = {3},
  pages = {033369},
  numpages = {20},
  year = {2020},
  month = {Sep},
  publisher = {American Physical Society},
  doi = {10.1103/PhysRevResearch.2.033369}
}

@article{King2022,
  title = {Coherent quantum annealing in a programmable 2,000 qubit Ising chain},
  author = {King, Andrew D. and Suzuki, Sei and Raymond, Jack and Zucca, Alex and Lanting, Trevor and Altomare, Fabio and Berkley, Andrew J. and Ejtemaee, Sara and Hoskinson, Emile and Huang, Shuiyuan and others},
  journal = {Nature Physics},
  volume = {18},
  pages = {1324--1328},
  year = {2022},
  doi = {10.1038/s41567-022-01741-6}
}

@article{King2023,
  title = {Quantum critical dynamics in a 5,000-qubit programmable spin glass},
  author = {King, Andrew D. and Raymond, Jack and Lanting, Trevor and Harris, Richard and Zucca, Alex and Altomare, Fabio and Berkley, Andrew J. and Boothby, Kelly and Ejtemaee, Sara and Enderud, Colin and others},
  journal = {Nature},
  volume = {617},
  number = {7959},
  pages = {61--66},
  year = {2023},
  doi = {10.1038/s41586-023-05867-2}
}

@article{zhang2024cyclic,
  title={Cyclic quantum annealing: searching for deep low-energy states in 5000-qubit spin glass},
  author={Zhang, Hao and Boothby, Kelly and Kamenev, Alex},
  journal={Scientific reports},
  volume={14},
  number={1},
  pages={30784},
  year={2024},
  publisher={Nature Publishing Group UK London}
}

@article{zhang2025computational,
  title={Computational complexity of three-dimensional Ising spin glass: Lessons from D-wave annealer},
  author={Zhang, Hao and Kamenev, Alex},
  journal={Physical Review Research},
  volume={7},
  number={3},
  pages={033098},
  year={2025},
  publisher={APS}
}

@article{arai2024,
  title = {Deep unfolded local quantum annealing},
  author = {Arai, Shunta and Takabe, Satoshi},
  journal = {Phys. Rev. Res.},
  volume = {6},
  issue = {4},
  pages = {043325},
  numpages = {9},
  year = {2024},
  month = {Dec},
  publisher = {American Physical Society},
  doi = {10.1103/PhysRevResearch.6.043325},
  url = {https://link.aps.org/doi/10.1103/PhysRevResearch.6.043325}
}

@article{xiao2026,
  title = {Globally guided simulated bifurcation for enhanced optimization},
  author = {Xiao, Zhijiao and Huang, Zujia and Qiu, Qijie and Liu, Yong-Qing and Zhuang, Jia-Pei and Yung, Man-Hong},
  journal = {Phys. Rev. Appl.},
  volume = {25},
  issue = {2},
  pages = {024014},
  numpages = {11},
  year = {2026},
  month = {Feb},
  publisher = {American Physical Society},
  doi = {10.1103/sdb2-stbs}
}

@misc{tao2026,
      title={Multi-Objective Optimization with Quantum-Annealing-Inspired Algorithms}, 
      author={Xian-Zhe Tao and Pavel Mosharev and Zhaofeng Su and Man-Hong Yung},
      year={2026},
      eprint={2604.26477},
      archivePrefix={arXiv},
      primaryClass={quant-ph},
      url={https://arxiv.org/abs/2604.26477}, 
}

@article{Amin2015,
  title = {Searching for quantum speedup in quasistatic quantum annealers},
  author = {Amin, Mohammad H.},
  journal = {Physical Review A},
  volume = {92},
  number = {5},
  pages = {052323},
  year = {2015},
  publisher = {American Physical Society},
  doi = {10.1103/PhysRevA.92.052323},
}

@misc{Shin2014,
      title={How "Quantum" is the D-Wave Machine?}, 
      author={Seung Woo Shin and Graeme Smith and John A. Smolin and Umesh Vazirani},
      year={2014},
      eprint={1401.7087},
      archivePrefix={arXiv},
      primaryClass={quant-ph},
      url={https://arxiv.org/abs/1401.7087}, 
}

@article{Subires2022,
  title = {Benchmarking quantum annealing dynamics: The spin-vector Langevin model},
  author = {Subires, David and G\'omez-Ruiz, Fernando J. and Ruiz-Garc\'{\i}a, Antonia and Alonso, Daniel and del Campo, Adolfo},
  journal = {Phys. Rev. Res.},
  volume = {4},
  issue = {2},
  pages = {023104},
  numpages = {10},
  year = {2022},
  month = {May},
  publisher = {American Physical Society},
  doi = {10.1103/PhysRevResearch.4.023104}
}

@article{Bowles2022,
  title = {Quadratic Unconstrained Binary Optimization via Quantum-Inspired Annealing},
  author = {Bowles, Joseph and Dauphin, Alexandre and Huembeli, Patrick and Martinez, Jos\'e and Ac\'{\i}n, Antonio},
  journal = {Phys. Rev. Appl.},
  volume = {18},
  issue = {3},
  pages = {034016},
  numpages = {10},
  year = {2022},
  month = {Sep},
  publisher = {American Physical Society},
  doi = {10.1103/PhysRevApplied.18.034016},
  url = {https://link.aps.org/doi/10.1103/PhysRevApplied.18.034016}
}

@article{Misra2023,
  title = {Mean-Field Approximate Optimization Algorithm},
  author = {Misra-Spieldenner, Aditi and Bode, Tim and Schuhmacher, Peter K. and Stollenwerk, Tobias and Bagrets, Dmitry and Wilhelm, Frank K.},
  journal = {PRX Quantum},
  volume = {4},
  issue = {3},
  pages = {030335},
  numpages = {19},
  year = {2023},
  month = {Sep},
  publisher = {American Physical Society},
  doi = {10.1103/PRXQuantum.4.030335},
  url = {https://link.aps.org/doi/10.1103/PRXQuantum.4.030335}
}

@article{Kibble1976,
  author  = {Kibble, T. W. B.},
  title   = {Topology of cosmic domains and strings},
  journal = {Journal of Physics A: Mathematical and General},
  volume  = {9},
  number  = {8},
  pages   = {1387--1398},
  year    = {1976},
  doi     = {10.1088/0305-4470/9/8/029}
}

@article{Zurek1985,
  author  = {Zurek, W. H.},
  title   = {Cosmological experiments in superfluid helium?},
  journal = {Nature},
  volume  = {317},
  pages   = {505--508},
  year    = {1985},
  doi     = {10.1038/317505a0}
}

@article{Zurek1996,
  author  = {Zurek, W. H.},
  title   = {Cosmological experiments in condensed matter systems},
  journal = {Physics Reports},
  volume  = {276},
  number  = {4},
  pages   = {177--221},
  year    = {1996},
  doi     = {10.1016/S0370-1573(96)00009-9}
}

@article{Dziarmaga2010,
  author  = {Dziarmaga, Jacek},
  title   = {Dynamics of a quantum phase transition and relaxation to a steady state},
  journal = {Advances in Physics},
  volume  = {59},
  number  = {6},
  pages   = {1063--1189},
  year    = {2010},
  doi     = {10.1080/00018732.2010.514702}
}

@article{Polkovnikov2011,
  author  = {Polkovnikov, Anatoli and Sengupta, Krishnendu and Silva, Alessandro and Vengalattore, Mukund},
  title   = {Colloquium: Nonequilibrium dynamics of closed interacting quantum systems},
  journal = {Reviews of Modern Physics},
  volume  = {83},
  number  = {3},
  pages   = {863--883},
  year    = {2011},
  doi     = {10.1103/RevModPhys.83.863}
}

@article{Dziarmaga2005,
  author  = {Dziarmaga, Jacek},
  title   = {Dynamics of a Quantum Phase Transition: Exact Solution of the Quantum Ising Model},
  journal = {Physical Review Letters},
  volume  = {95},
  number  = {24},
  pages   = {245701},
  year    = {2005},
  doi     = {10.1103/PhysRevLett.95.245701}
}

@article{Zurek2005,
  author  = {Zurek, Wojciech H. and Dorner, Uwe and Zoller, Peter},
  title   = {Dynamics of a Quantum Phase Transition},
  journal = {Physical Review Letters},
  volume  = {95},
  number  = {10},
  pages   = {105701},
  year    = {2005},
  doi     = {10.1103/PhysRevLett.95.105701}
}

@article{LagunaZurek1998,
  title = {Critical dynamics of symmetry breaking: Quenches, dissipation, and cosmology},
  author = {Laguna, Pablo and Zurek, Wojciech Hubert},
  journal = {Phys. Rev. D},
  volume = {58},
  issue = {8},
  pages = {085021},
  numpages = {4},
  year = {1998},
  month = {Sep},
  publisher = {American Physical Society},
  doi = {10.1103/PhysRevD.58.085021},
  url = {https://link.aps.org/doi/10.1103/PhysRevD.58.085021}
}

@article{Lipkin1965,
  author  = {Lipkin, H. J. and Meshkov, N. and Glick, A. J.},
  title   = {Validity of many-body approximation methods for a solvable model: (I). Exact solutions and perturbation theory},
  journal = {Nuclear Physics},
  volume  = {62},
  number  = {2},
  pages   = {188--198},
  year    = {1965},
  doi     = {10.1016/0029-5582(65)90862-X}
}

@article{Botet1982,
  author  = {Botet, R. and Jullien, R. and Pfeuty, P.},
  title   = {Size Scaling for Infinitely Coordinated Systems},
  journal = {Physical Review Letters},
  volume  = {49},
  number  = {7},
  pages   = {478--481},
  year    = {1982},
  doi     = {10.1103/PhysRevLett.49.478}
}

@article{Xie2024,
  title = {Quantum Hopfield model with dilute memories},
  author = {Xie, Rongfeng and Kamenev, Alex},
  journal = {Phys. Rev. A},
  volume = {110},
  issue = {3},
  pages = {032418},
  numpages = {10},
  year = {2024},
  month = {Sep},
  publisher = {American Physical Society},
  doi = {10.1103/PhysRevA.110.032418},
  url = {https://link.aps.org/doi/10.1103/PhysRevA.110.032418}
}

@misc{Li2026,
      title={Kibble--Zurek Scaling in the Dicke Model at Mesoscopic Scales}, 
      author={Haowei Li and Hanteng Wang},
      year={2026},
      eprint={2608.20067},
      archivePrefix={arXiv},
      primaryClass={cond-mat.quant-gas},
      url={https://arxiv.org/abs/2608.20067}, 
}

@article{Tyagi2025,
  title = {Asymmetry amplification by a nonadiabatic passage through a critical point},
  author = {Tyagi, Bhavay and Suzuki, Fumika and Chernyak, Vladimir A. and Sinitsyn, Nikolai A.},
  journal = {Phys. Rev. A},
  volume = {111},
  issue = {3},
  pages = {032205},
  numpages = {6},
  year = {2025},
  month = {Mar},
  publisher = {American Physical Society},
  doi = {10.1103/PhysRevA.111.032205},
  url = {https://link.aps.org/doi/10.1103/PhysRevA.111.032205}
}

@article{Derrida1986,
  title = {Metastable states of a spin glass chain at 0 temperature},
  author = {Derrida, B. and Gardner, E.},
  journal = {Journal de Physique},
  volume = {47},
  number = {6},
  pages = {959--965},
  year = {1986},
  doi = {10.1051/jphys:01986004706095900}
}

@article{Igloi2005,
  title = {Strong disorder {RG} approach of random systems},
  author = {Igl{\'o}i, Ferenc and Monthus, C{\'e}cile},
  journal = {Physics Reports},
  volume = {412},
  number = {5--6},
  pages = {277--431},
  year = {2005},
  doi = {10.1016/j.physrep.2005.02.006},
}

@article{Vojta2006,
  title = {Rare region effects at classical, quantum and nonequilibrium phase transitions},
  author = {Vojta, Thomas},
  journal = {Journal of Physics A: Mathematical and General},
  volume = {39},
  number = {22},
  pages = {R143--R205},
  year = {2006},
  doi = {10.1088/0305-4470/39/22/R01}
}

@article{Ma1979,
  title = {Random Antiferromagnetic Chain},
  author = {Ma, Shang-keng and Dasgupta, Chandan and Hu, Chin-kun},
  journal = {Phys. Rev. Lett.},
  volume = {43},
  issue = {19},
  pages = {1434--1437},
  numpages = {0},
  year = {1979},
  month = {Nov},
  publisher = {American Physical Society},
  doi = {10.1103/PhysRevLett.43.1434}
}

@article{Dasgupta1980,
  title = {Low-temperature properties of the random Heisenberg antiferromagnetic chain},
  author = {Dasgupta, Chandan and Ma, Shang-keng},
  journal = {Phys. Rev. B},
  volume = {22},
  issue = {3},
  pages = {1305--1319},
  numpages = {0},
  year = {1980},
  month = {Aug},
  publisher = {American Physical Society},
  doi = {10.1103/PhysRevB.22.1305}
}

@article{Fisher1995,
  title = {Critical behavior of random transverse-field Ising spin chains},
  author = {Fisher, Daniel S.},
  journal = {Phys. Rev. B},
  volume = {51},
  issue = {10},
  pages = {6411--6461},
  numpages = {0},
  year = {1995},
  month = {Mar},
  publisher = {American Physical Society},
  doi = {10.1103/PhysRevB.51.6411},
  url = {https://link.aps.org/doi/10.1103/PhysRevB.51.6411}
}

@article{Igloi2001,
  title = {Griffiths-McCoy Singularities in Random Quantum Spin Chains: Exact Results through Renormalization},
  author = {Igl\'oi, Ferenc and Juh\'asz, R\'obert and Lajk\'o, P\'eter},
  journal = {Phys. Rev. Lett.},
  volume = {86},
  issue = {7},
  pages = {1343--1346},
  numpages = {0},
  year = {2001},
  month = {Feb},
  publisher = {American Physical Society},
  doi = {10.1103/PhysRevLett.86.1343},
  url = {https://link.aps.org/doi/10.1103/PhysRevLett.86.1343}
}

@article{Igloi2002,
  title = {Exact renormalization of the random transverse-field Ising spin chain in the strongly ordered and strongly disordered Griffiths phases},
  author = {Igl\'oi, Ferenc},
  journal = {Phys. Rev. B},
  volume = {65},
  issue = {6},
  pages = {064416},
  numpages = {11},
  year = {2002},
  month = {Jan},
  publisher = {American Physical Society},
  doi = {10.1103/PhysRevB.65.064416},
  url = {https://link.aps.org/doi/10.1103/PhysRevB.65.064416}
}

@article{Dziarmaga2006,
  title = {Dynamics of a quantum phase transition in the random {I}sing model: Logarithmic dependence of the defect density on the transition rate},
  author = {Dziarmaga, Jacek},
  journal = {Phys. Rev. B},
  volume = {74},
  issue = {6},
  pages = {064416},
  numpages = {5},
  year = {2006},
  month = {Aug},
  publisher = {American Physical Society},
  doi = {10.1103/PhysRevB.74.064416},
  url = {https://link.aps.org/doi/10.1103/PhysRevB.74.064416}
}

@article{Caneva2007,
  title = {Adiabatic quantum dynamics of a random Ising chain across its quantum critical point},
  author = {Caneva, Tommaso and Fazio, Rosario and Santoro, Giuseppe E.},
  journal = {Phys. Rev. B},
  volume = {76},
  issue = {14},
  pages = {144427},
  numpages = {9},
  year = {2007},
  month = {Oct},
  publisher = {American Physical Society},
  doi = {10.1103/PhysRevB.76.144427},
  url = {https://link.aps.org/doi/10.1103/PhysRevB.76.144427}
}

@article{Suzuki2009,
doi = {10.1088/1742-5468/2009/03/P03032},
url = {https://doi.org/10.1088/1742-5468/2009/03/P03032},
year = {2009},
month = {mar},
publisher = {},
volume = {2009},
number = {03},
pages = {P03032},
author = {Suzuki, Sei},
title = {Cooling dynamics of pure and random Ising chains},
journal = {Journal of Statistical Mechanics: Theory and Experiment}
}

@article{EdwardsAnderson1975,
  title = {Theory of spin glasses},
  author = {Edwards, S. F. and Anderson, P. W.},
  journal = {Journal of Physics F: Metal Physics},
  volume = {5},
  number = {5},
  pages = {965--974},
  year = {1975}
}

@article{Bieche1980,
  title = {On the ground states of the frustration model of a spin glass by a matching method of graph theory},
  author = {Bieche, I. and Maynard, R. and Rammal, R. and Uhry, J. P.},
  journal = {Journal of Physics A: Mathematical and General},
  volume = {13},
  number = {8},
  pages = {2553--2576},
  year = {1980},
  doi = {10.1088/0305-4470/13/8/005}
}

@article{Barahona1982,
  title = {On the computational complexity of Ising spin glass models},
  author = {Barahona, F.},
  journal = {Journal of Physics A: Mathematical and General},
  volume = {15},
  number = {10},
  pages = {3241--3253},
  year = {1982},
  doi = {10.1088/0305-4470/15/10/028}
}

@article{Pardella2008,
  title = {Exact ground states of large two-dimensional planar Ising spin glasses},
  author = {Pardella, G. and Liers, F.},
  journal = {Physical Review E},
  volume = {78},
  number = {5},
  pages = {056705},
  year = {2008},
  doi = {10.1103/PhysRevE.78.056705}
}

@article{Rubin2017,
  title = {Dual time scales in simulated annealing of a two-dimensional Ising spin glass},
  author = {Rubin, Shanon J. and Xu, Na and Sandvik, Anders W.},
  journal = {Phys. Rev. E},
  volume = {95},
  issue = {5},
  pages = {052133},
  numpages = {12},
  year = {2017},
  month = {May},
  publisher = {American Physical Society},
  doi = {10.1103/PhysRevE.95.052133},
  url = {https://link.aps.org/doi/10.1103/PhysRevE.95.052133}
}

@article{Xu2017,
  title = {Dynamic scaling in the two-dimensional Ising spin glass with normal-distributed couplings},
  author = {Xu, Na and Wu, Kai-Hsin and Rubin, Shanon J. and Kao, Ying-Jer and Sandvik, Anders W.},
  journal = {Phys. Rev. E},
  volume = {96},
  issue = {5},
  pages = {052102},
  numpages = {9},
  year = {2017},
  month = {Nov},
  publisher = {American Physical Society},
  doi = {10.1103/PhysRevE.96.052102},
  url = {https://link.aps.org/doi/10.1103/PhysRevE.96.052102}
}

@article{Huse1986,
  title = {Residual Energies after Slow Cooling of Disordered Systems},
  author = {Huse, David A. and Fisher, Daniel S.},
  journal = {Phys. Rev. Lett.},
  volume = {57},
  issue = {17},
  pages = {2203--2206},
  numpages = {0},
  year = {1986},
  month = {Oct},
  publisher = {American Physical Society},
  doi = {10.1103/PhysRevLett.57.2203},
  url = {https://link.aps.org/doi/10.1103/PhysRevLett.57.2203}
}

@article{Weinberg2020,
  title = {Scaling and Diabatic Effects in Quantum Annealing with a D-Wave Device},
  author = {Weinberg, Phillip and Tylutki, Marek and R{\"o}nkk{\"o}, Jami M. and Westerholm, Jan and {\AA{}}str{\"o}m, Jan A. and Manninen, Pekka and T{\"o}rm{\"a}, P{\"a}ivi and Sandvik, Anders W.},
  journal = {Phys. Rev. Lett.},
  volume = {124},
  issue = {9},
  pages = {090502},
  numpages = {6},
  year = {2020},
  month = {Mar},
  publisher = {American Physical Society},
  doi = {10.1103/PhysRevLett.124.090502},
  url = {https://link.aps.org/doi/10.1103/PhysRevLett.124.090502}
}

@article{Schmitt2022,
  author = {Schmitt, Markus and Rams, Marek M. and Dziarmaga, Jacek and Heyl, Markus and Zurek, Wojciech H.},
  title = {Quantum phase transition dynamics in the two-dimensional transverse-field {Ising} model},
  journal = {Science Advances}, volume = {8}, number = {37}, pages = {eabl6850}, year = {2022},
  doi = {10.1126/sciadv.abl6850}, eprint = {2106.09046}, archivePrefix = {arXiv}, primaryClass = {cond-mat.str-el}
}

@misc{WeinbergXuSandvik2025,
  author = {Weinberg, Phillip and Xu, Na and Sandvik, Anders W.},
  title = {Defects and their Time Scales in Quantum and Classical Annealing of the Two-Dimensional {Ising} Model},
  year = {2025}, eprint = {2507.09273}, archivePrefix = {arXiv}, primaryClass = {quant-ph},
  url = {https://arxiv.org/abs/2507.09273}
}

@misc{Naik2026,
      title={Real-time Dynamics in 3D for up to 1000 Qubits with Neural Quantum States: Quenches and the Quantum Kibble--Zurek Mechanism}, 
      author={Vighnesh Dattatraya Naik and Zheng-Hang Sun and Markus Heyl},
      year={2026},
      eprint={2604.05032},
      archivePrefix={arXiv},
      primaryClass={quant-ph},
      url={https://arxiv.org/abs/2604.05032}
}
\end{document}